\documentclass[preprint,showpacs,eqsecnum,aps]{revtex4}
\usepackage{graphicx}
\usepackage{epsfig}
\usepackage{multirow}
\begin{document}

\title{Application of the sextic oscillator with centrifugal barrier and the spheroidal equation for some X(5) candidate nuclei}

\author{A. A. Raduta$^{a,b}$, P. Buganu$^{a}$}

\address{$^{a)}$ Department of Theoretical Physics, Institute of Physics and Nuclear Engineering, P.O.B. MG-6, RO-077125, Romania}

\address{$^{b)}$Academy of Romanian Scientists, 54 Splaiul Independentei, Bucharest 050094, Romania}

\begin{abstract}
The eigenvalue equation associated to the Bohr-Mottelson Hamiltonian is considered in the intrinsic reference frame and amended by replacing the harmonic oscillator potential in the $\beta$ variable with a sextic oscillator potential with centrifugal barrier plus a periodic potential for the $\gamma$ variable. After the separation of variables,  the $\beta$ equation is quasi-exactly solved, while the solutions for the $\gamma$ equation  are just the angular spheroidal functions. An anharmonic transition operator is used to determine the reduced E2 transition probabilities. The formalism is conventionally called the Sextic and Spheroidal Approach (SSA) and applied for several X(5) candidate nuclei: $^{176,178,180,188,190}$Os, $^{150}$Nd, $^{170}$W, $^{156}$Dy, $^{166,168}$Hf. The SSA predictions are in good agreement with the experimental data of the mentioned nuclei. The comparison of the SSA results with those yielded by other models, such as X(5) \cite{Iache9}, Infinite Square Well (ISW) \cite{Raduta}, and Davidson (D) like potential \cite{Raduta} for the $\beta$, otherwise keeping the spheroidal functions for the $\gamma$ , and the Coherent State Model (CSM) \cite{Rad1,Rad2,Rad3,Rad4,RaSa,Rad5} respectively, suggests that SSA represents a good approach to describe nuclei achieving the critical point of the U(5)$\rightarrow$SU(3) shape phase transition.
\end{abstract}

\maketitle

\renewcommand{\theequation}{1.\arabic{equation}}
\setcounter{equation}{0}
\section{Introduction}
\label{sec:level1}
Since the liquid drop model was developed \cite{Bohr}, the
quadrupole shape coordinates  were widely used  by both
phenomenological and microscopic formalisms to describe the basic properties of
nuclear systems. Based on these coordinates, one defines quadrupole
boson operators in terms of which model Hamiltonians and transition operators
are defined. Since the original spherical harmonic liquid drop model was able
to describe only a small amount of data for spherical nuclei, several
improvements have been added. Thus, the Bohr-Mottelson model was generalized by
Faessler and Greiner \cite{GrFa}
in order to describe the small oscillations around a deformed shape which
results in obtaining a flexible model, called vibration rotation model,
suitable for the description of deformed nuclei. Later on \cite{Gneus} this picture was
extended by including anharmonicities as low order invariant  polynomials in the
quadrupole coordinates. With a suitable choice of the parameters involved in the model
Hamiltonian the equipotential
energy surface may exhibit several types of minima \cite{Hess} like spherical,
deformed prolate, deformed oblate, deformed triaxial, etc.
 To each equilibrium shape, specific properties for excitation energies and electromagnetic transition
 probabilities show up. Due to this reason,
one customarily says that static values of intrinsic coordinates determine a phase for the
nuclear system. The boson description with a complex anharmonic Hamiltonian makes use of a large number of structure parameters which are
to be fitted.  A smaller number of parameters is used by the
coherent state model (CSM) \cite{Rad1} which uses a restricted collective space generated through angular momentum projection by three deformed orthogonal functions of coherent type. The model is able to describe in a realistic fashion transitional and well deformed nuclei of various shapes including states of high and very high angular momentum. Various extensions to include other degrees of freedom like isospin
 \cite{Rad2}, single particle \cite{Rad3} or octupole \cite{Rad4,RaSa} degrees of freedom have been formulated \cite{Rad5}.
  
It has been noticed that a given nuclear shape may be
associated with a certain symmetry. Hence, its properties may be described with
the help of the irreducible representation of the respective symmetry group.
Thus, the gamma unstable nuclei can be described by the $O(6)$ symmetry
\cite{Jean}, the gamma-rigid triaxial rotor by the $D2$ symmetry
\cite{Filip},
the symmetric rotor by the $SU(3)$ symmetry and the spherical vibrator by the
$U(5)$ symmetry.
Thus, even in the 50's, the   symmetry properties have been greatly appreciated.
 However, a big push forward was brought by the interacting boson
 approximation
(IBA) \cite{Iache,Iache1}, which succeeded to describe the basic properties of a large number of
nuclei in terms of the symmetries associated to a system of quadrupole (d) and
monopole (s) bosons  which  generate the $U(6)$ algebra of the IBA. The three
limiting symmetries $U(5)$, $O(6)$ and $SU(3)$ mentioned above in the context of the collective model are  also dynamic symmetries
for $U(6)$. Moreover, for each of these symmetries a specific group reduction
chain  provides the quantum numbers characterizing the states, which are suitable
  for a certain region of nuclei. Besides  the virtue of unifying the group
  theoretical descriptions of nuclei exhibiting different symmetries,
  the procedure defines very simple reference pictures for the limiting cases. For nuclei
 lying close to the region characterized by a certain symmetry,
 the perturbative corrections are to be included.

 In Refs. \cite{Gino,Diep}, it has been proved
that on the $U(5)-O(6)$ transition leg there exists a critical point
for a second order phase transition while the
$U(5)-SU(3)$ leg has a first order phase transition. Actually, the first order phase transition takes place not only on the mentioned leg of the Casten's triangle but covers all te interior of the triangle up to the second order \cite{McCutchan1}. Examples of such nuclei,  falling insider the triangle, are the $Os$ isotopes \cite{McCutchan2}.   .

Recently, Iachello \cite{Iache2,Iache9} pointed out that these critical points
correspond to
distinct symmetries, namely $E(5)$ and $X(5)$, respectively. For the critical value
of an
ordering parameter, energies are given by the zeros of a Bessel function of half integer and irrational indices, respectively.

The description of low lying states in terms of Bessel functions was used first by Jean and Willet \cite{Jean}, but the interesting feature saying that this is a critical picture in a phase transition and defines a new symmetry, was indeed advanced first in Ref.\cite{Iache2}. 

Representatives for
the two symmetries have been experimentally identified. To give an example, the relevant data for $^{134}$Ba \cite{Zam} and $^{152}$Sm \cite{Zam1}
suggest that they are close to the $E(5)$ and $X(5)$ symmetries, respectively.
Another candidate for E(5) symmetry, is $^{102}$Pd \cite{Zam2,Da-li}.
A systematic search for $E(5)$ behavior in nuclei has been reported in Ref.\cite{Clar}.

In Ref.\cite{Rad05} we advanced the hypothesis that the critical point in a phase transition is state dependent. We tested this with a hybrid model for $^{134}$Ba and $^{104}$Ru. Similar property of the phase transition was investigated  in the context of a schematic two level model in
Ref. \cite{Leyv,Gil}. A rigorous analysis of the the characteristics of excited state quantum phase transitions is performed in Ref. \cite{Capr}. 

The departure from the $\gamma$ unstable picture has been treated by several authors \cite{Bona} whose contributions are reviewed by Fortunato in Ref.\cite{Fortunato7}. The difficulty in treating the $\gamma$ degree of freedom consists in the fact that this variable is coupled to the rotation variables. A full solution for the Bohr-Mottelson Hamiltonian including an explicit treatment of $\gamma$ deformation variable can be found in Refs.\cite{Ghe78,Rad78,Moshi,Corri,Marge}. Therein, we treated separately also the $\gamma$ unstable and the rotor Hamiltonian. A more complete study of the rotor Hamiltonian and the distinct phases associated to a tilted moving rotor is given in Ref. \cite{Rad98}.

The treatment of the $\gamma$ variable becomes even more complicated when we add to the liquid drop Hamiltonian a potential depending on $\beta$ and $\gamma$ at a time.
 To simplify the starting problem related to the inclusion of the $\gamma$ variable one uses  model potentials which are sums of a beta and a $\gamma$ depending potentials. In this way the nice feature for the beta variable to be decoupled from the remaining 4 variables, specific to the harmonic liquid drop, is preserved. Further the potential in $\gamma$ is expanded either around to $\gamma=0$ or around $\gamma=\frac{\pi}{6}$.
In the first case if only the singular term is retained one obtains the infinite square well model described by Bessel functions in gamma. If the $\gamma^2$ term is added to this term, the Laguerre functions are the eigenstates of the approximated gamma depending Hamiltonian, which results in defining the functions characterizing the X(5) approach.

The drawback of these approximation consists in that the resulting $\gamma$ depending functions are not periodic as the starting Hamiltonian is. Moreover, they are orthonormalized on unbound intervals although the underlying equation was derived under the condition of $|\gamma|$ small. The scalar product for  the space of the resulting functions is not defined based on the measure $|\sin3\gamma| d\gamma$ as  happens in the liquid drop model. Under these circumstances it happens that the approximated Hamiltonian in $\gamma$ looses its hermiticity.

In some earlier publications \cite{Rad07,Raduta} we proposed a scheme where the gamma variable is described by a solvable Hamiltonian whose eigenstates are spheroidal functions which are periodic. Here we give details about the calculations and describe some new numerical applications. Moreover, the formalism was completed by treating the $\beta$ variable by a Schr\"{o}dinger equation associated to the Davidson's potential. Alternatively we considered the equation for a five dimensional square well potential. We have shown that the new treatment of the gamma variable removes the drawbacks mentioned above and moreover brings a substantial improvement of the numerical analysis.

Here we keep the description of the gamma variable by spheroidal functions and use a new
potential for the beta variable which seems to be more suitable for a realistic description of 
more complex spectra. We call this approach as Sextic and Spheroidal Approach ($SSA$). The potential is that of a sextic oscillator plus a centrifugal term which
leads to a quasi-exactly solvable model. The resulting formalism will be applied to 10 nuclei which were not included in our previous descriptions and moreover are suspected to be good candidate  for exhibiting $X(5)$ features having the ratio of excitation energies of the ground band members  $4^+$ and $2^+$ close to the value of 2.9.
The results of our calculations are compared with those obtained through other methods  such as ISW, D and CSM.

The goals presented in the previous paragraph will be developed according to the following plan. In Section II  the main ingredients of the theoretical models 
$X(5)$, $ISW$, $D$ and $SSA$ will be briefly presented. The $CSM$ is separately described in Section III. Numerical results are given and commented in Section IV, while the final conclusions are drown in Section V.  

\renewcommand{\theequation}{2.\arabic{equation}}
\setcounter{equation}{0}
\section{The separation of variables and solutions}
\label{sec:level2}
In order to describe  the critical nuclei  of the  U(5)$-$SU(3) shape phase transition, we resort the Bohr-Mottelson Hamiltonian with a potential  depending  on both  the $\beta$ and $\gamma$ variables:
\begin{equation}
H\psi(\beta,\gamma,\Omega)=E\psi(\beta,\gamma,\Omega),
\label{eigen}
\end{equation}
where
\begin{equation}
H=-\frac{\hbar^2}{2B}\left[\frac{1}{\beta^4}\frac{\partial }{\partial \beta}
\beta^4 \frac{\partial }{\partial \beta}+\frac{1}{\beta^2\sin {3\gamma} }
\frac{\partial}{\partial \gamma}\sin{3\gamma}\frac{\partial}{\partial \gamma}
-\frac{1}{4\beta^2}\sum_{k=1}^{3}\frac{\hat{Q}_k^2}{\sin^2(\gamma-\frac{2\pi}{3}k)}
\right]+V(\beta,\gamma).
\label{Has}
\end{equation}
Here, $\beta$ and $\gamma$ are the intrinsic deformation variables,  $\Omega$  denotes the Euler angles $\theta_{1}$, $\theta_{2}$ and $\theta_{3}$, $\hat{Q}_{k}$ are the angular momentum components  in the intrinsic reference frame, while $B$ is the so called mass parameter.

\subsection{The separation of variables}

To achieve the separation of variables in  Eq. (\ref{eigen}), some approximations are necessary. Choosing the potential energy in the form \cite{Jean,Fortunato7}
\begin{equation}
V(\beta,\gamma)=V_{1}(\beta)+\frac{V_{2}(\gamma)}{\beta^{2}},
\end{equation}
the $\beta$ variable is separated from the $\gamma$ and the Euler angles $\Omega$, which are still coupled due to the rotational term:
\begin{equation}
W=\frac{1}{4}\sum_{k=1}^{3}\frac{\hat{Q}_{k}^{2}}{\sin^{2}\left(\gamma-\frac{2\pi}{3}k\right)}.
\end{equation}
Further, the $\gamma$ is separated from the Euler angles by using the second order power expansion of the rotational term around the equilibrium value $\gamma_{0}=0^{0}$ (see Eq. (B.5) from Ref. \cite{Raduta}):
\begin{equation}
W\approx\frac{1}{3}\hat{Q}^{2}+\left(\frac{1}{4\sin^{2}\gamma}-\frac{1}{3}\right)\hat{Q}_{3}^{2}+\frac{2}{2\sqrt{3}}(\hat{Q}_{2}^{2}-\hat{Q}_{1}^{2})\gamma+\frac{2}{3}(\hat{Q}^{2}-\hat{Q}_{3}^{2})\gamma^{2}+\mathcal{O}(\gamma^{3}),
\end{equation}
and then averaging the result with the  Wigner function $D_{M,K}^{(L)}$:
\begin{equation}
\langle W\rangle=\frac{1}{3}L(L+1)+\left(\frac{1}{4\sin^{2}\gamma}-\frac{1}{3}\right)K^{2}+\frac{2}{3}[L(L+1)-K^{2}]\gamma^{2}.
\label{eqrot}
\end{equation}
The term $L(L+1)/3$  multiplied by $1/\beta^{2}$  is transferred to the  equation for $\beta$,
\begin{equation}
\left[-\frac{1}{\beta^{4}}\frac{\partial}{\partial\beta}\beta^{4}\frac{\partial}{\partial\beta}+\frac{L(L+1)}{3\beta^{2}}+v_{1}(\beta)\right]f(\beta)=\varepsilon_{\beta}f(\beta),
\label{eqbeta}
\end{equation}
while the  sum of remaining terms, denoted with $\tilde{V}(\gamma,L,K)$, are kept in the equation for $\gamma$.
\begin{equation}
\left[-\frac{1}{\sin3\gamma}\frac{\partial}{\partial\gamma}\sin3\gamma\frac{\partial}{\partial\gamma}+\tilde{V}(\gamma,L,K)+v_{2}(\gamma)\right]\eta(\gamma)=\tilde{\varepsilon}_{\gamma}\eta(\gamma).
\label{eqgamma}
\end{equation}
In Eqs. (\ref{eqbeta}) and (\ref{eqgamma}) the following notations were used:
\begin{equation}
v_{1}(\beta)=\frac{2B}{\hbar^{2}}V_{1}(\beta),\hspace{0.5cm}v_{2}(\gamma)=\frac{2B}{\hbar^{2}}V_{2}(\gamma),\hspace{0.5cm}\varepsilon_{\beta}=\frac{2B}{\hbar^{2}}E_{\beta},\hspace{0.5cm}\tilde{\varepsilon}_{\gamma}=\langle\beta^{2}\rangle\frac{2B}{\hbar^{2}}E_{\gamma}.
\label{enreduse}
\end{equation}
Eqs. (\ref{eqbeta}) and (\ref{eqgamma}) are to  be separately solved  and finally the full solution of Eq. (\ref{eigen}) is obtained by combining the contributions coming from each variable.
In what follows we shall give the necessary details for solving the above mentioned equations.

\subsection{Solutions of the $\beta$ equation}

Solutions of the $\beta$ equation, corresponding to  different potentials, were considered by several authors \cite{Fortunato7,Cejnar}. Here, we mention only three of them, namely  the infinite square well, the Davidson and the sextic potentials. Details about how to solve the $\beta$ equation  for these potentials can be found in Refs. \cite{Raduta,RaBu011}.

\subsubsection{The infinite square well potential}
The solution of the $\beta$ equation with an infinite square well potential, having the expression
\begin{equation}
v_{1}(\beta)=\Bigg\{{{0,\,\,\beta\leq\beta_{\omega}} \atop {\infty,\,\,\beta>\beta_{\omega}}},
\label{udeb}
\end{equation}
was first time given in Ref. \cite{Jean} and then in Refs. \cite{Iache2,Iache9} for 
E(5) and X(5) models. The $\beta$ wave functions are written in terms of the Bessel functions of half integer \cite{Iache2} and irrational indices \cite{Iache9}, respectively. The solution for X(5) is:
\begin{equation}
f_{s,L}(\beta)=C_{s,L}\beta^{-\frac{3}{2}}J_{\nu}\left(\frac{x_{s,L}}{\beta_{\omega}}\beta\right),\;\;\;\nu=\sqrt{\frac{L(L+1)}{3}+\frac{9}{4}},\;\;s=1,2,3,... .
\end{equation}
Here, $C_{s,L}$ is the normalization factor, which is determined from the condition:
\begin{equation}
\int_{0}^{\beta_{\omega}}(f_{s,L}(\beta))^{2}\beta^{4}d\beta=1.
\end{equation}
The corresponding eigenvalues are given in terms of  the Bessel zeros $x_{s,L}$:
\begin{equation}
E_{\beta}(s,L)=\frac{\hbar^{2}}{2B}\left(\frac{x_{s,L}}{\beta_{\omega}}\right)^2.
\label{eigenISW}
\end{equation}

\subsubsection{The Davidson potential}
Choosing in Eq. (\ref{eqbeta}) a Davidson potential \cite{Davi} of the form
\begin{equation}
v_{1}(\beta )=\beta ^{2}+\frac{\beta_0^4}{\beta ^2},
\label{Davi}
\end{equation}
 solutions  are the generalized Laguerre polynomials:
\begin{equation}
f_{n_{\beta},m_{\beta}}(\beta)=\sqrt{\frac{2n_{\beta}!}{\Gamma(n_{\beta}+m_{\beta}+1)}}L_{n_{\beta}}^{m_{\beta}}(\beta^{2})\beta^{m_{\beta}-\frac{3}{2}}e^{-\frac{\beta^{2}}{2}},\;\;m_{\beta}=\sqrt{\frac{L(L+1)}{3}+\frac{9}{4}+\beta_{0}^{4}}.
\label{solDav}
\end{equation}
The wave functions, $f_{n_{\beta},m_{\beta}}(\beta)$ are normalized to unity with the integration measure $\beta^{4}d\beta$.
Energies have the following expression:
\begin{equation}
E_{\beta}(n_{\beta},L)=\frac{\hbar^{2}}{2B}\left(2n_{\beta}+1+\sqrt{\frac{L(L+1)}{3}+\frac{9}{4}+\beta_{0}^{4}}\right),\;\;n_{\beta}=0,1,2,...,\;\;n_{\beta}=s-1.
\label{eigenDav}
\end{equation}

\subsubsection{The sextic oscillator potential with a centrifugal barrier}
The solution of the $\beta$ equation with a sextic potential, for critical nuclei of the  U(5)$\rightarrow$SU(3) shape phase transition, was obtained by taking into consideration the solution of the Schr\"{o}dinger equation with a sextic  potential given in Ref. \cite{Ush} and  applied to the E(5) like nuclei  in Ref. \cite{Levai} and to the triaxial nuclei in Ref.\cite{RaBu011}. 

In order, to reduce the $\beta$ equation to the Schr\"{o}dinger equation with a sextic  potential \cite{Ush}, we rewrite the averaged rotational term, given by  Eq. (\ref{eqrot}), in the following form:
\begin{equation}
\langle W\rangle=[L(L+1)-2]+\left[2-\frac{2}{3}L(L+1)\right]+\left(\frac{1}{4\sin^{2}\gamma}-\frac{1}{3}\right)K^{2}+\frac{2}{3}[L(L+1)-K^{2}]\gamma^{2}.
\label{rotorsextic}
\end{equation}
As already mentioned, the first term of the above equation is added to the $\beta$ equation, while the other terms remain in the $\gamma$ equation. Making the substitution $f(\beta)=\beta^{-2}\varphi(\beta)$ we have:
\begin{equation}
\left[-\frac{\partial^{2}}{\partial\beta^{2}}+\frac{L(L+1)}{\beta^{2}}+v_{1}(\beta)\right]\varphi(\beta)=\varepsilon_{\beta}\varphi(\beta).
\label{newbetaeq}
\end{equation}
The sextic potential is chosen such that to obtain the description from Ref.\cite{RaBu011}:
\begin{equation}
\hspace{0.4cm}v_{1}^{\pm}(\beta)=(b^{2}-4ac^{\pm})\beta^{2}+2ab\beta^{4}+a^{2}\beta^{6}+u_{0}^{\pm},\hspace{0.2cm}c^{\pm}=\frac{L}{2}+\frac{5}{4}+M.
\label{v1debeta}
\end{equation}
Here, $c$ is a constant which has two different values, one for $L$ even and other for $L$ odd:
\begin{equation}
(M,L):(k,0);(k-1,2);(k-2,4);(k-3,6)...\Rightarrow c=k+\frac{5}{4}\equiv c^{+}\hspace{0.2cm}(L\mbox{-even}),
\end{equation}
\begin{equation}
(M,L):(k,1);(k-1,3);(k-2,5);(k-3,7)...\Rightarrow c=k+\frac{7}{4}\equiv c^{-}\hspace{0.2cm}(L\mbox{-odd}).
\end{equation}
The constants $u_{0}^{\pm}$ are fixed such that the potential for $L$ odd has the same minimum energy as the potential for $L$ even. The solutions of Eq. (\ref{newbetaeq}), with the potential given by the Eq. (\ref{v1debeta}), are
\begin{equation}
\varphi_{n_{\beta},L}^{(M)}(\beta)=N_{n_{\beta},L}P_{n_{\beta},L}^{(M)}(\beta^{2})\beta^{L+1}e^{-\frac{a}{4}\beta^{4}-\frac{b}{2}\beta^{2}},\hspace{0.2cm}n_{\beta}=0,1,2,...M,
\label{sexticbf}
\end{equation}
where $N_{n_{\beta},L}$ are the normalization factor, while $P_{n_{\beta},L}^{(M)}(\beta^{2})$ are polynomials in $x^{2}$ of $n_{\beta}$ order. The corresponding excitation energy is:
\begin{equation}
E_{\beta}(n_{\beta},L)=\frac{\hbar^{2}}{2B}\left[b(2L+3)+\lambda_{n_{\beta}}^{(M)}(L)+u_{0}^{\pm}\right],\hspace{0.2cm}n_{\beta}=0,1,2,...,M,
\label{energybe}
\end{equation}
where $\lambda_{n_{\beta}}^{(M)}=\varepsilon_{\beta}-u_{0}^{\pm}-4bs$ is the eigenvalue of the equation:
\begin{equation}
\left[-\left(\frac{\partial^{2}}{\partial\beta^{2}}+\frac{4s-1}{\beta}\frac{\partial}{\partial \beta}\right)+2b\beta\frac{\partial}{\partial\beta}+2a\beta^{2}\left(\beta\frac{\partial}{\partial \beta}-2M\right)\right]P_{n_{\beta},L}^{(M)}(\beta^{2})=\lambda_{n_{\beta}}^{(M)}P_{n_{\beta},L}^{(M)}(\beta^{2}).
\label{Qoper}
\end{equation}

\subsection{Solutions of the $\gamma$ equation}

\subsubsection{The X(5) model}
Within the X(5) model \cite{Iache9}, devoted to  the description of the critical point in the phase transition  $SU(5)\to SU(3)$, the potential is a sum of an infinite square well in the $\beta$ variable and a harmonic oscillator in the $\gamma$ variable. For the rotational term and the other terms of the $\gamma$ equation, the first order Taylor  expansion around $\gamma_{0}=0^{0}$ is considered, which results in obtaining for the $\gamma$ variable the radial equation of a two dimensional oscillator with the solution
\begin{equation}
\eta_{n_{\gamma},K}(\gamma)=C_{n,K}\gamma^{|K/2|}e^{-(3a)\gamma^{2}/2}L_{n}^{|K|}(3a\gamma^{2}),\;\;n=\left(\frac{n_{\gamma}-|K|}{2}\right),
\end{equation}
where $L_{n}^{|K|}$ are the generalized Laguerre polynomials. The eigenvalue of the $\gamma$ equation has the following expression:
\begin{equation}
\varepsilon_{\gamma}=\frac{3a}{\sqrt{\langle\beta^{2}\rangle}}(n_{\gamma}+1)-\frac{(K/2)^{2}}{\langle\beta^{2}\rangle}\frac{4}{3},
\end{equation}
where $a$ is a parameter characterizing the oscillator potential in the $\gamma$ variable. The total energy and wave function is obtain by combining the results of all variables:
\begin{equation}
E(s,L,n_{\gamma},K)=E_{0}+B_1(x_{s,L})^{2}+An_{\gamma}+CK^{2},
\label{x5energy}
\end{equation}
\begin{equation}
\Psi(\beta,\gamma,\Omega)=\frac{1}{\sqrt{2(1+\delta_{K.0})}}f_{s,L}(\beta)\left[\eta_{n_{\gamma},K}(\gamma)D_{M,K}^{L}(\Omega)+(-1)^{L+K}\eta_{n_{\gamma},-K}(\gamma)D_{M,-K}^{L}(\Omega)\right].
\end{equation}
If the total energy (\ref{x5energy}) is normalized to the energy of the ground state, we will have for the ground band and for the first beta band the expression
\begin{equation}
E(s,L,0,0)-E(1,0,0,0)=B_1(x_{s,L}^{2}-x_{1,0}^{2}),\;\;s=1,2;\;\;L=0,2,4,6,...
\end{equation}
while for the first $\gamma$ band
\begin{equation}
E(s,L,1,2)-E(1,0,0,0)=B_1(x_{1,L}^{2}-x_{1,0}^{2})+A+4C,\;\;L=2,3,4,5,... .
\end{equation}
One notes that the parameters $A$ and $C$ give contribution only to the $\gamma$ band energies, and that these two parameters can be replaced with only one parameter, for example $X=A+4C$. The total energy for the ground band and for the first $\beta$ and $\gamma$ bands, normalized to the energy of the ground state, can be written in the form:
\begin{equation}
E(s,L,n_{\gamma},K)-E(1,0,0,0)=B_1(x_{s,L}^{2}-x_{1,0}^{2})+\delta_{K,2}X.
\end{equation}
Further the parameters $B_1$ and $X$ will be fitted by the least square procedure for each considered nucleus.
\subsubsection{The ISW model}
Within the ISW model, employed in the present paper, the $\beta$ equation is treated as in 
 the X(5) model, using an infinite square well (ISW), while the $\gamma$ equation is reduced to a spheroidal equation. The ISW model was proposed, by one of the authors (A.A.R) and his collaborators in Ref. \cite{Rad07} and subsequently with more details and applications in Ref.
\cite{Raduta}. Here, only the solutions will be presented. The potential $v_{2}(\gamma)$ was chosen such that a minimum in $\gamma=0^{0}$ is achieved:
\begin{equation}
v_{2}(\gamma)=u_{1}\cos3\gamma+u_{2}\cos^{2}3\gamma.
\label{potgamma}
\end{equation}
This potential is renormalized by a contribution coming from the $\gamma$ rotational term and consequently an effective reduced potential for the $\gamma$ variable results
\begin{equation}
\tilde{v}_{2}(\gamma)=u_{1}\cos3\gamma+u_{2}\cos^{2}3\gamma+\frac{9}{4\sin^{3}3\gamma},
\label{effpot}
\end{equation}
whose minima are shifted with respect to the $v_2(\gamma)$ minima.
This can be viewed as the reduced potential of:
\begin{equation}
\tilde{V}_2=\frac{\hbar^2}{2B}\tilde{v}_2.
\end{equation}
Performing a second order expansion in $\sin3\gamma$ of $v_{2}(\gamma)$ and of the terms originating from the rotational term i.e., $\frac{9}{4\sin^{3}3\gamma}$, and then
making the change of variable $x=\cos3\gamma$ in Eq. (\ref{eqgamma}) we obtain the equation for the spheroidal functions \cite{Raduta}:
\begin{equation}
\left[(1-x^{2})\frac{\partial^{2}}{\partial x^{2}}-2x\frac{\partial}{\partial x}+\lambda_{m_{\gamma},n_{\gamma}}-c^{2}x^{2}-\frac{m_{\gamma}^{2}}{1-x^{2}}\right]S_{m_{\gamma},n_{\gamma}}(x)=0,
\label{spheroidalequation}
\end{equation}
where
\begin{eqnarray}
&&\lambda_{m_{\gamma},n_{\gamma}}=\frac{1}{9}\left[\tilde{\varepsilon}_{\gamma}-\frac{u_{1}}{2}-\frac{11}{27}D+\frac{1}{3}L(L+1)\right],\nonumber\\
&&c^{2}=\frac{1}{9}\left(\frac{u_{1}}{2}+u_{2}-\frac{2}{27}D\right),\nonumber\\
&&m_{\gamma}=\frac{K}{2},\;\;D=L(L+1)-K^{2}-2.
\label{eigenvaluespher}
\end{eqnarray}
From Eq. (\ref{eigenvaluespher}) we can determine the eigenvalue of the $\gamma$ equation:
\begin{equation}
E_{\gamma}(n_{\gamma},m_{\gamma},L,K)=\frac{1}{\langle\beta^{2}\rangle}\frac{\hbar^{2}}{2B}\left(9\lambda_{m_{\gamma},n_{\gamma}}(c)+\frac{u_{1}}{2}+\frac{11}{27}D-\frac{L(L+1)}{3}\right).
\label{eigenvaluegamma}
\end{equation}
In Eq. (\ref{eigenvaluegamma}), the term $u_{1}/2$ is washed out when the total energy is normalized to the ground state energy, which results in getting  the $\gamma$ eigenvalue depending on the sum of the $\gamma$ potential parameters, due to the term $c^{2}$. Hence, in some cases we can set one parameter to be equal to zero, for example $u_{2}$, and consequently fit only $u_{1}$. The $\gamma$ functions are normalized to unity with the integration measure $|\sin3\gamma| d\gamma$ as the Bohr-Mottelson model requires:
\begin{equation}
\frac{3(2n_{\gamma}+1)(n_{\gamma}-m_{\gamma})!}{2(n_{\gamma}+m_{\gamma})!}\int_{0}^{\frac{\pi}{3}}|S_{m_{\gamma},n_{\gamma}}(\cos3\gamma)|^{2}|\sin3\gamma|d\gamma=1.
\end{equation}
The total energy is obtained by summing the contributions coming from the $\beta$ (\ref{eigenISW}) and the $\gamma$ (\ref{eigenvaluegamma}) equations:
\begin{equation}
E(s,n_{\gamma},m_{\gamma},L,K)=B_1x_{s,L}^{2}+F\left[9\lambda_{m_{\gamma},n_{\gamma}}(c)+\frac{u_{1}}{2}+\frac{11}{27}D-\frac{L(L+1)}{3}\right],
\end{equation}
where the following notations were introduced:
\begin{equation}
B_1=\frac{1}{\beta_{\omega}^{2}}\frac{\hbar^{2}}{2B},\;\;F=\frac{1}{\langle\beta^{2}\rangle}\frac{\hbar^{2}}{2B}.
\end{equation}
The total wave function is:
\begin{equation}
\Psi(\beta,\gamma,\Omega)=C_{s,L}C_{n_{\gamma},m_{\gamma}}C_{L,K}\beta^{-\frac{3}{2}}J_{\nu}\left(\frac{x_{s,L}}{\beta_{\omega}}\beta\right)S_{m_{\gamma},n_{\gamma}}(\cos3\gamma)\left[D_{M,K}^{L}(\Omega)+(-1)^{L}D_{M,-K}^{L}(\Omega)\right],
\end{equation}
where with $C_{n_{\gamma},m_{\gamma}}$ was denoted the normalization factor of the $\gamma$ function, while $C_{L,K}$ is the normalization factor of the Wigner function:
\begin{equation}
C_{L,K}=\sqrt{\frac{2L+1}{16\pi^{2}(1+\delta_{K,0})}}.
\end{equation}

\subsubsection{The D model}
The $D$ model was proposed by the present authors and collaborators in Ref. \cite{Raduta} and differs from the ISW model by that the infinite square well potential for the $\beta$ variable is replaced with the Davidson potential (\ref{Davi}). Hence, the total energy of the system is 
obtained by adding the energy of the $\beta$ equation with Davidson potential given by Eq. (\ref{eigenDav}) and the energy of the $\gamma$ equation (\ref{eigenvaluegamma}):
\begin{eqnarray}
E(n_{\beta},n_{\gamma},m_{\gamma},L,K)&=&E\left(2n_{\beta}+1+\sqrt{\frac{L(L+1)}{3}+\frac{9}{4}+\beta_{0}^{4}}\right)\nonumber\\
&+&F\left[9\lambda_{m_{\beta},n_{\gamma}}(c)+\frac{u_{1}}{2}+\frac{11}{27}D-\frac{L(L+1)}{3}\right],
\end{eqnarray}
where $E=\hbar^{2}/2B$.
The total wave function has the expression:
\begin{equation}
\Psi(\beta,\gamma,\Omega)=C_{n_{\beta},L}C_{n_{\gamma},m_{\gamma}}C_{L,K}f_{n_{\beta},L}(\beta)S_{m_{\gamma},n_{\gamma}}(\cos3\gamma)\left[D_{M,K}^{L}(\Omega)+(-1)^{L}D_{M,-K}^{L}(\Omega)\right],
\end{equation}
where with $C_{n_{\beta},L}$ is the normalization factor of $f_{n_{\beta},L}(\beta)$ given by the Eq. (\ref{solDav}).

\subsubsection{The present approach}
In the present approach, called conventionally the Sextic and Spheroidal Approach (SSA), a sextic potential (\ref{v1debeta}) for the $\beta$ variable is considered, while for the $\gamma$ variable  a periodic potential (\ref{potgamma}) with a minimum at $\gamma_{0}=0^{0}$. The $\beta$ equation is quasi-exactly solved, having the solutions given by the Eqs. (\ref{sexticbf},\ref{energybe}), while the $\gamma$ equation is reduced to the spheroidal equation (\ref{spheroidalequation}) with:
\begin{eqnarray}
&&\lambda_{m_{\gamma},n_{\gamma}}=\frac{1}{9}\left[\tilde{\varepsilon}_{\gamma}-\frac{u_{1}}{2}-\frac{11}{27}D+\frac{1}{3}L(L+1)\right]+\frac{2L(L+1)}{27},\nonumber\\
&&c^{2}=\frac{1}{9}\left(\frac{u_{1}}{2}+u_{2}-\frac{2}{27}D\right),\nonumber\\
&&m_{\gamma}=\frac{K}{2},\;\;D=L(L+1)-K^{2}-2.
\label{eigensphersextic}
\end{eqnarray}
In Eq. (\ref{eigensphersextic}), the term $2L(L+1)/3$ multiplied with $1/9$ comes from the rotational term (\ref{rotorsextic}). The expression for the total energy of the system is obtained by using the Eqs. (\ref{energybe},\ref{eigensphersextic}):
\begin{equation}
E(n_{\beta},n_{\gamma},m_{\gamma},L,K)=E\left[b(2L+3)+\lambda_{n_{\beta}}^{(M)}+u_{0}^{\pm}\right]+F\left[9\lambda_{m_{\beta},n_{\gamma}}(c)+\frac{u_{1}}{2}+\frac{11}{27}D-L(L+1)\right].
\end{equation}
The corresponding wave function, is:
\begin{equation}
\Psi(\beta,\gamma,\Omega)=N_{n_{\beta},L}C_{n_{\gamma},m_{\gamma}}C_{L,K}\beta^{-2}\varphi_{n_{\beta},L}(\beta)S_{m_{\gamma},n_{\gamma}}(\cos3\gamma)\left[D_{M,K}^{L}(\Omega)+(-1)^{L}D_{M,-K}^{L}(\Omega)\right],
\end{equation}
where $\varphi_{n_{\beta},L}(\beta)$ is given by  Eq. (\ref{sexticbf}).
\subsection{E2 transition probabilities}
The reduced E2 transition probabilities are determined by:
\begin{equation}
B(E2;L_{i}\rightarrow L_{f})=|\langle L_{i}||T_{2}^{(E2)}||L_{f}\rangle|^{2},
\label{redtrans}
\end{equation}
where the Rose's convention \cite{Rose} was used. For the ISW, D and SSA models, in 
Eq. (\ref{redtrans}), an anharmonic transition operator is used:
\begin{eqnarray}
T^{(E2)}_{2\mu}&=&t_1\beta \left[\cos\gamma D^2_{\mu 0}(\Omega)+\frac{\sin\gamma}{\sqrt{2}}(D^2_{\mu 2}(\Omega)+D^2_{\mu, -2}(\Omega))\right]+\nonumber\\
         & &t_2\sqrt{\frac{2}{7}}\beta^2 \left[-\cos2\gamma D^2_{\mu 0}(\Omega)+\frac{\sin2\gamma}{\sqrt{2}}(D^2_{\mu 2}(\Omega)+D^2_{\mu, -2}(\Omega))\right].
\label{transop}
\end{eqnarray}
The parameters $t_{1}$ and $t_{2}$ will be determined by the least squares method. For the X(5) model, in the limit of $\gamma-$small, only the harmonic part of the transition operator (\ref{transop}) is used:
\begin{equation}
T_{2\mu,X(5)}^{(E2)}=t\beta D_{\mu0}^{2}(\Omega)+t\beta\frac{\gamma}{\sqrt{2}}(D_{\mu2}^{2}(\Omega)+D_{\mu,2}^{2}(\Omega)).
\label{transX5}
\end{equation}
The first term of the Eq. (\ref{transX5}) gives contributions only to $\triangle K=0$ transitions, while the second term to $\triangle K=2$ transitions. For $\triangle K=0$ transitions, the matrix element of the $\gamma$ variable is reduced to the orthogonality condition, while for $\triangle K=2$ the $\gamma$ matrix element can be considered as an intrinsic transition matrix element. Finally, the reduced transition probabilities  will depend on two parameters \cite{Bij}. Here, we will denote these two parameters with $t$ for $\triangle K=0$ transitions and $t'$ for $\triangle K=2$ transitions, respectively.
\renewcommand{\theequation}{3.\arabic{equation}}
\setcounter{equation}{0}
\section{The coherent state model}
\label{sec:level3}
CSM defines \cite{Rad1} first a restricted collective space whose vectors  are
model states of ground, $\beta$ and $\gamma$ bands. In choosing these states we
were guided by some experimental information which results in formulating a set of criteria to be fulfilled by the searched states.

All these restrictions required are fulfilled by the following set of three deformed quadrupole boson states:
\begin{equation}
\psi_g=e^{[d(b^{\dagger}_0-b_0)]}|0\rangle\equiv T|0\rangle,~
\psi_{\gamma}=\Omega^{\dagger}_{\gamma,2}\psi_g,~
\psi_{\beta}=\Omega^{\dagger}_{\beta}\psi_g.
\label{psigbga}
\end{equation}
where the excitation operators for $\beta$ and $\gamma$ bands are defined by:
\begin{equation}
\Omega^{\dagger}_{\gamma,2}=(b^{\dagger}b^{\dagger})_{2,2}+d\sqrt{\frac{2}{7}}
b^{\dagger}_{2,2},\;\;\Omega^{\dagger}_{\beta}=(b^{\dagger}b^{\dagger}b^{\dagger})_0 +\frac{3d}{\sqrt{14}}
(b^{\dagger}b^{\dagger})_0 -\frac{d^3}{\sqrt{70}}.
\label{omegabe}
\end{equation}
Here, $d$ is a real parameter simulating the nuclear deformation. From the three deformed states one generates through projection, three sets of
mutually orthogonal states
\begin{equation}
\varphi^i_{JM}=N^i_JP^J_{M0}\psi_i, i=g,\beta,\gamma,
\label{projfi}
\end{equation}
where $P^J_{MK}$ denotes the projection operator:
\begin{equation}
P^J_{MK}=\frac{2J+1}{8\pi^2}\int {D^{J^*}_{MK}\hat{R}(\Omega)d\Omega},
\label{projop}
\end{equation}
$N^i_{J}$  the normalization factors and $D^J_{MK}$ the rotation matrix elements. The rotation operator corresponding to the Euler angles $\Omega$ is denoted by $\hat{R}(\Omega)$.
It was proved that the deformed and projected states contain the salient features of the major collective bands.
Since we attempt to set up a very simple model we relay on the experimental feature saying that the $\beta$  band is largely decoupled from the ground as well as from the $\gamma$ bands and choose a model Hamiltonian  whose matrix elements
between  beta states and  states belonging either to the ground or to the gamma band are all equal to zero.
The simplest Hamiltonian obeying this restriction is
\begin{equation}
H=A_1(22\hat{N}+5\Omega^{\dag}_{\beta'}\Omega_{\beta'})+A_2\hat{J}^2
+A_3\Omega^{\dagger}_{\beta}\Omega_{\beta},
\label{hascsm}
\end{equation}
where $\hat{N}$ is the boson number, $\hat{J}^2$-angular momentum squared and $
\Omega^{\dagger}_{\beta'}$ denotes:
\begin{equation}
\Omega^{\dagger}_{\beta'}=(b^{\dagger}b^{\dagger})_{00}-\frac{d^2}{\sqrt{5}}.
\end{equation}

Higher order terms in boson operators can be added to the Hamiltonian $H$ without altering the decoupling condition for the beta band. An example of this kind is the correction:
\begin{equation}
\Delta H=A_4(\Omega^{\dagger}_{\beta}\Omega^2_{\beta^{\prime}}+h.c.)+
A_5\Omega^{\dagger 2}_{\beta^{\prime}}\Omega^2_{\beta^{\prime}}.
\label{deltahascsm}
\end{equation}

The energies for beta band as well as for the gamma band states of odd
angular momentum are described as average values of H (\ref{hascsm}), or $H+\Delta H$ on $\varphi^{\beta}_{JM}$
and $\varphi^{\gamma}_{JM}$ (J-odd), respectively. As for the energies for the ground band and
those of gamma band states  with even angular momentum, they are obtained by diagonalizing
a 2x2 matrix for each J.

The quadrupole transition operator is considered to be a sum of a linear  term in bosons and one which is quadratic in the quadrupole bosons:
\begin{equation}
Q_{2\mu}=q_1(b^{\dag}_{2\mu}+(-)^{\mu}b_{2,-\mu})+
q_2((b^{\dag}_2b^{\dag}_2)_{2\mu}+(b_{\tilde{2}}b_{\tilde{2}})_{2\mu}) +q_3(b^{\dagger}_2b_{\tilde{2}})_{2\mu} .
\label{q2anh}
\end{equation}
Note that if $q_3=2q_2$ the quadrupole transition operator can be obtained from the quadrupole transition operator expressed in terms of the collective quadruple coordinates $\alpha_{2\mu}$:
\begin{equation}
Q_{2\mu}=Q_1\alpha_{2\mu}+Q^{\prime}_1(\alpha_2\alpha_2)_{2\mu}.
\end{equation}
The anharmonic term in the above expression can be obtained by expanding the deformed mean field around the spherical equilibrium shape \cite{RCD76,Maruhn} of the nuclear surface. For the near vibrational regime the interband matrix elements of the $q_3$ term is vanishing within the CSM \cite{Rad1}. Moreover, a transition operator depending on two free parameters seems to be  suitable for describing the E2 transition probabilities in several regions of the nuclides chart \cite{RaBud012}.

Using the Rose convention \cite{Rose}, the reduced probability for the E2 transition 
$J^+_i\to J^+_f$ can be expressed as:
\begin{equation}
B(E2;J^+_i\to J^+_f)=\left(\langle J^+_i||Q_2||J^+_f\rangle\right)^2
\end{equation}
Three specific features of CSM are worth to be mentioned:

a) The model states are generated through projection from a coherent
state and two excitations of that through simple  polynomial boson operators.
Thus,
it is expected that the projected states may account for the semiclassical
behavior of the nuclear system staying in a state of high spin.

b) The states are infinite series of bosons and thus highly deformed
states can be described.

c) The model Hamiltonian is not commuting with the boson number operator and
because of this property a basis generated from a coherent state is expected
to be most suitable.

The CSM has been successfully applied to several nuclei
exhibiting various equilibrium shapes which according to the IBA (Interacting Boson Approximation) classification,
exhibit the  SO(6), SU(5) and SU(3) symmetries, respectively.
Several improvements of CSM has been proposed by considering additional
degrees of freedom
like isospin \cite{Rad2}, quasiparticle \cite{Rad3} or collective octupole coordinates
\cite{Rad4,RaSa}. CSM has been also used to describe some nonaxial nuclei \cite{RadUve} and the results were compared with those obtained with the Rotation-Vibration Model \cite{GrFa}.
A review of the CSM achievements is found in Ref. \cite{Rad5}. The terms involved in the model Hamiltonians used in by CSM \cite{Rad1} and its generalized version \cite{Rad2} have microscopic counterparts as shown in \cite{AAR76} and \cite{AAR00}, respectively.

\renewcommand{\theequation}{4.\arabic{equation}}
\setcounter{equation}{0}
\section{Numerical results}
\label{sec:level4}

\subsection{Parameters}
The parameters which define the energies and the E2 transitions probabilities of the models X(5), ISW, D, SSA and CSM, where fitted by the least squares method for ten nuclei: $^{176,178,180,188,190}$Os, $^{150}$Nd, $^{156}$Dy, $^{166,168}$Hf and $^{170}$W. In the least square procedure all experimental energies were considered. The resulting values are those given in the Tables I-V. For the first three and the last three nuclei from Table I, the parameter $t'$ cannot be determined since the corresponding term from the transition operator does not contribute to the intraband decays.

Some parameters  vary by a large amount from one isotope to another but the relative variation is small. For example in the case of $Os$ isotopes the parameters could be interpolated by  smooth curves. One parameter is falling aside namely those of $^{188}$Os, which seems to achieve the critical point of the shape transition, i.e. exhibits a $X(5)$ behavior. 

We note that the parameter $F$ involves the average value $\langle \beta^2\rangle$ which, in principle, is an angular momentum dependent quantity. Therefore the differential equation in $\gamma$ should be iteratively solved, at each step the inserted average value being calculated with the wave function provided in the previous step. When the convergence of the process is met, one keeps the average value for the chosen angular momentum. Here, $\langle \beta^2\rangle$ was taken constant. Whether this hypothesis is valid or not can be posterity checked. To this goal we represented in Fig. 1 the average $\langle \beta^2\rangle$ for each of the models $ISW$, $D$ and $SSA$. We notice that the average value is only slightly depending on $J$ and that is especially true for $ISW$ and $SSA$. If the limit of  $\langle \beta^2\rangle$ when the convergence of the iterations mentioned above is reached, depends on $J$ like the averages shown in Fig. 1, one could say that keeping $\langle \beta^2\rangle$ constant one ignores a slight decrease of energy with angular momentum.

\begin{table}[h!]
\caption{The fitted values of the parameters involved in the expressions of the energies and transition probabilities of the X(5) model are given for each considered nucleus.}
\vspace{0.3cm}
\begin{tabular}{|c|c|c|c|c|c|c|c|c|c|c|}
  \hline
  X(5) & $^{176}$Os & $^{178}$Os & $^{180}$Os & $^{188}$Os & $^{190}$Os & $^{150}$Nd & $^{156}$Dy & $^{166}$Hf & $^{168}$Hf & $^{170}$W \\
   \hline
   $B_1$ [keV] & 18.08 & 18.13 & 18.79 & 25.56 & 26.92 & 17.77 & 17.02 & 23.46 & 20.14 & 20.68 \\
   \hline
   X [keV]& 822.28 & 818.68 & 880.10 & 452.71 & 438.53 & 966.50 & 950.46 & 698.15 & 770.26 & 799.14 \\
   \hline
   t $[W.u.]^{1/2}$& 1.29 & 1.22& 0.84 & 0.86 & 0.76 & 1.03 & 1.19 & 0.99 & 1.19 & 0.89 \\
   \hline
   t$'$ $[W.u.]^{1/2}$ & - & - & - & 0.92 & 1.19 & 0.49 & 0.81 & - & - & - \\
  \hline
\end{tabular}
\end{table}

\begin{table}[h!]
\caption{The same as in Table I, but for the ISW model.}
\vspace{0.3cm}
\begin{tabular}{|c|c|c|c|c|c|c|c|c|c|c|}
  \hline
   ISW & $^{176}$Os & $^{178}$Os & $^{180}$Os & $^{188}$Os & $^{190}$Os & $^{150}$Nd & $^{156}$Dy & $^{166}$Hf & $^{168}$Hf & $^{170}$W \\
   \hline
   $B_1$ [keV] & 14.30 & 14.54 & 13.21 & 25.50 & 21.83 & 14.68 & 11.43 & 23.31 & 19.12 & 14.87 \\
   \hline
   F [keV]& 24.24 & 23.19 & 44.66 & 0.69 & 36.73 & 28.88 & 45.99 & 1.69 & 11.30 & 41.12 \\
   \hline
   u$_{1}$& -159.24 & -168.08 & -36.729 & -25000 & -4999.35 & -152.35 & -12.55 & -10000 & -385.35 & -44.36 \\
   \hline
   u$_{2}$& 0 & 0 & 0 & 0 & 2560.22 & 0 & 0 & 0 & 0 & 0 \\
   \hline
   t$_{1}$ $[W.u.]^{1/2}$& -52.91 & 473.53 & 3302.3 & 503.11 & 419.67 & 538.99 & 591.54 & 1881.39 & 1197.94 & 1827.11 \\
   \hline
   t$_{2}$ $[W.u.]^{1/2}$& -4305.14 & -1323.6 & 14304.2 & -241.19 & -48.09 & -387.08 & -468.57 & 8242.45 & 2702.98 & 6436.57 \\
  \hline
\end{tabular}
\end{table}

\begin{table}[h!]
\caption{The same as in Table I, but for the D model.}
\vspace{0.3cm}
\begin{tabular}{|c|c|c|c|c|c|c|c|c|c|c|}
  \hline
  D & $^{176}$Os & $^{178}$Os & $^{180}$Os & $^{188}$Os & $^{190}$Os & $^{150}$Nd & $^{156}$Dy & $^{166}$Hf & $^{168}$Hf & $^{170}$W \\
   \hline
   E [keV] & 316.34 & 317.31 & 334.32 & 559.76 & 462.44 & 369.50 & 324.08 & 532.22 & 463.88 & 379.93 \\
   \hline
   F [keV]& 38.41 & 37.33 & 39.01 & 28.48 & 42.45 & 26.48 & 33.11 & 11.87 & 25.87 & 37.72 \\
   \hline
   $\beta_{0}$ & 1.64 & 1.56 & 1.61 & 1.98 & 1.64 & 1.71 & 1.45 & 1.79 & 2.02 & 1.63 \\
   \hline
   u$_{1}$& -55.48 & -57.20 & -52.40 & -7.70 & -4098.61 & -168.78 & -58.16 & -320.01 & -130.49 & -54.50 \\
   \hline
   u$_{2}$& 0 & 0 & 0 & 0 & 2167.18 & 0 & 0 & 0 & 0 & 0 \\
   \hline
   t$_{1}$ $[W.u.]^{1/2}$& 197.92 & 264.47 & 758.41 & 126.88 & 126.70 & 154.70 & 191.28 & 448.76 & 329.01 & 411.63 \\
   \hline
   t$_{2}$ $[W.u.]^{1/2}$& -25.31 & 78.30 & 931.21 & -17.09 & -3.92 & -25.31 & -13.46 & 430.42 & 193.06 & 363.66 \\
\hline
\end{tabular}
\end{table}

\begin{table}[h!]
\caption{The same as in Table I, but for the SSA model.}
\vspace{0.3cm}
\begin{tabular}{|c|c|c|c|c|c|c|c|c|c|c|}
  \hline
  SSA & $^{176}$Os & $^{178}$Os & $^{180}$Os & $^{188}$Os & $^{190}$Os & $^{150}$Nd & $^{156}$Dy & $^{166}$Hf & $^{168}$Hf & $^{170}$W \\
   \hline
   E [keV] & 0.99 & 0.46 & 1.46 & 2.53 & 5.29 & 0.75 & 0.91 & 1.82 & 0.54 & 0.31 \\
   \hline
   F [keV]& 2.67 & 3.12 & 1.69 & 11.31 & 5.55 & 3.87 & 1.93 & 15.97 & 1.99 & 2.84 \\
   \hline
   a & 951.49 & 4466.56 & 600.70 & 644.98 & 111.79 & 2636.48 & 1248.40 & 1205.13 & 7897.62 & 13197.99 \\
   \hline
   b & 126 & 279 & 50 & 27 & 15.8 & 88 & 87 & 46 & 32 & 341 \\
   \hline
   u$_{1}$& -5607.45 & -4048.06 & -15000 & -215.19 & -452.74 & -3877.84 & -10000 & -224.90 & -9980.01 & -4585.44 \\
   \hline
   u$_{2}$& 0 & 0 & 0 & 0 & 0 & 0 & 0 & 0 & 0 & 0 \\
   \hline
   t$_{1}$ $[W.u.]^{1/2}$& 376.70 & 2260.6 & 8541.32 & 1033.43 & 675.12 & 1754.26 & 1882.91 & 4759.23 & 3463.05 & 8901.59 \\
   \hline
   t$_{2}$ $[W.u.]^{1/2}$ & -32619.3 & -22343.8 & 117781 & -1022.41 & 32.73 & -6698.41 & -4846.17 & 46113.3 & 15247.9 & 200989\\
  \hline
\end{tabular}
\end{table}

\begin{table}[h!]
\caption{The same as in Table I, but for the CSM model.}
\vspace{0.3cm}
\begin{tabular}{|c|c|c|c|c|c|c|c|c|c|c|}
  \hline
   CSM & $^{176}$Os & $^{178}$Os & $^{180}$Os & $^{188}$Os & $^{190}$Os & $^{150}$Nd & $^{156}$Dy & $^{166}$Hf & $^{168}$Hf & $^{170}$W \\
   \hline
   A$_{1}$ [keV] &17.03  &17.26   &16.51  &10.25&9.063  &19.219&15.45&14.87  &16.04  &16.19  \\
   \hline
   A$_{2}$ [keV]&4.33    &4.32    &5.19  & 14.40 &15.68  &3.467&5.2  &7.13  &6.40  &6.018  \\
   \hline
   A$_{3}$ [keV]&-395.96  &-240.13&-7.39 &101.362  &6.84  &-658.299&-559.913&-5.04&-61.47  &-186.946  \\
   \hline
   A$_{4}$ [keV]&-275.24  &-158.87&13.83 &0.0  &0.0  & -491.884 &-398.775&0.0&-36.48  &-124.55  \\
   \hline
   A$_{5}$ [keV]&-4.93 &30.76     &80.01 &0.0  &0.0  &-438.394  &-32.15&0.0  &0.0  & 0.0 \\
   \hline
   d            &2.33  & 2.36     &2.26  &2.35 &2.05&2.42  & 2.1 &2.08  &2.43  & 2.14 \\
   \hline
   q$_{1}$ $[W.u.]^{1/2}$&0.411  &0.246    & 0.86 & 0.409 &0.229  &0.527  &1.112&0.158  &0.211  &-0.217  \\
   \hline
   q$_{2}$ $[W.u.]^{1/2}$&-3.698  &-3.862 & 6.99 &0.785  &1.213 &-4.916  &-9.474&-5.075  &-3.936  & -5.602 \\
  \hline
   q$_{3}$ $[W.u.]^{1/2}$& 0.0    & 0.0   & 0.0 &-5.222  & -9.395 &6.344&19.576&0.0  & 0.0 &0.0  \\
   \hline
\end{tabular}
\end{table}
With the parameters listed above the potentials in the variables $\beta$ and $\gamma$ and the wave functions describing the low lying states from the ground, beta and gamma bands respectively, are  represented for four nuclei in Figs. II-V. Analyzing these figures, several features can be noticed. The $\beta$ potential has a deformed minimum located at a deformation which differs from one nucleus to another. The wave functions in $\beta$ for $0^+_{g}$ and $2^+_{\gamma}$  are almost identical and have only one maximum and no node while the band for $0^+_{\beta}$ has one node, one maximum and one minimum. The maximum of the $|\phi|^2$ distribution for the three states represented in the quoted figures is achieved in a point which is close to the potential minimum. If $|\phi|^2$ is multiplied with the integration measure over $\beta$ the probability distribution has a maximum closer to the potential minimum. The state $0^+_{\beta}$ is characterized by two maxima for the probability distribution of the beta variable. This feature reflects the specific structure of the excitation operator of this state, from the ground state i.e., $n_{\beta}=1$. The behavior of the wave functions in the variable $\gamma$ is mainly determined by the discontinuity for $\gamma =0$ and $\gamma=\frac{\pi}{3}$. The potential has two minima, one well pronounced near the first wall and one very flat close to the $\gamma=\frac{\pi}{3}$ discontinuity. Due to this structure the wave function 
describing a state in the ground bad has two maxima located above the mentioned minima. The state $2^+_{\gamma}$ heading the gamma band has an additional maximum. 

\begin{figure}[h!]
\begin{center}
\includegraphics[width=0.6\textwidth]{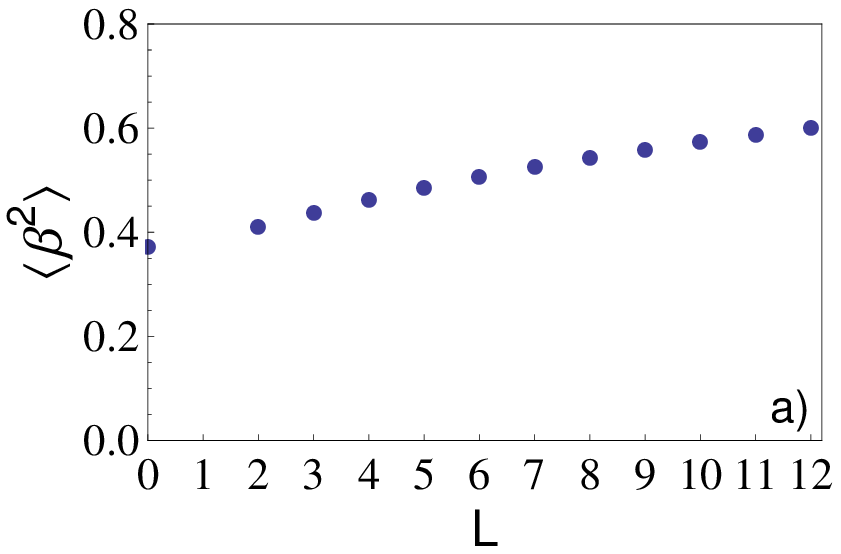}
\includegraphics[width=0.6\textwidth]{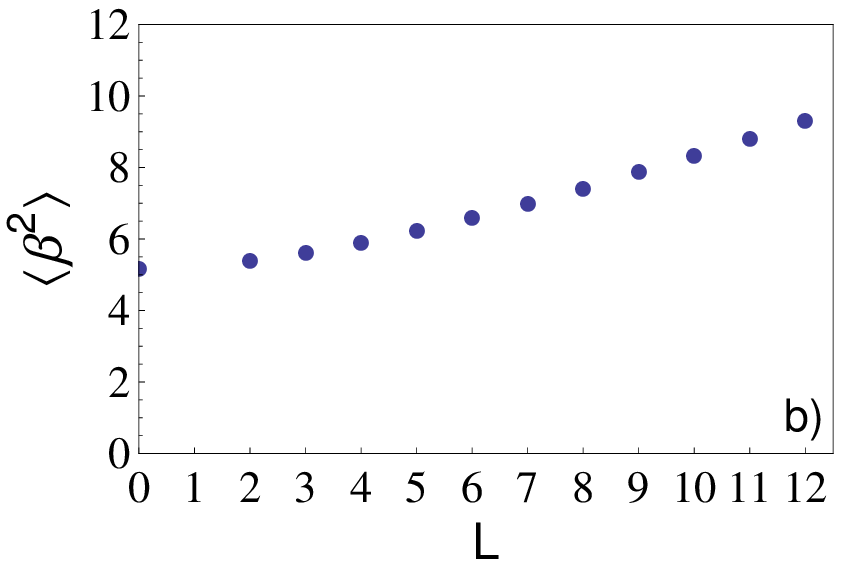}
\includegraphics[width=0.6\textwidth]{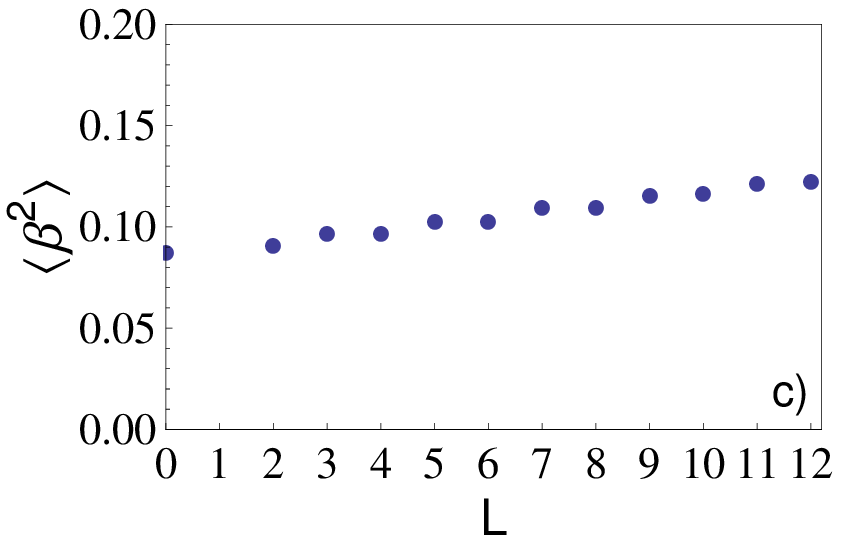}
\end{center}
\caption{The average values of $\beta^2$ vs. the  angular momentum calculated within the ISW (panel a)), the D (panel b)) and the SSA (panel c)) models.}
\label{Fig. 0}    
\end{figure}

\begin{figure}[h!]
\begin{center}
\includegraphics[width=0.8\textwidth]{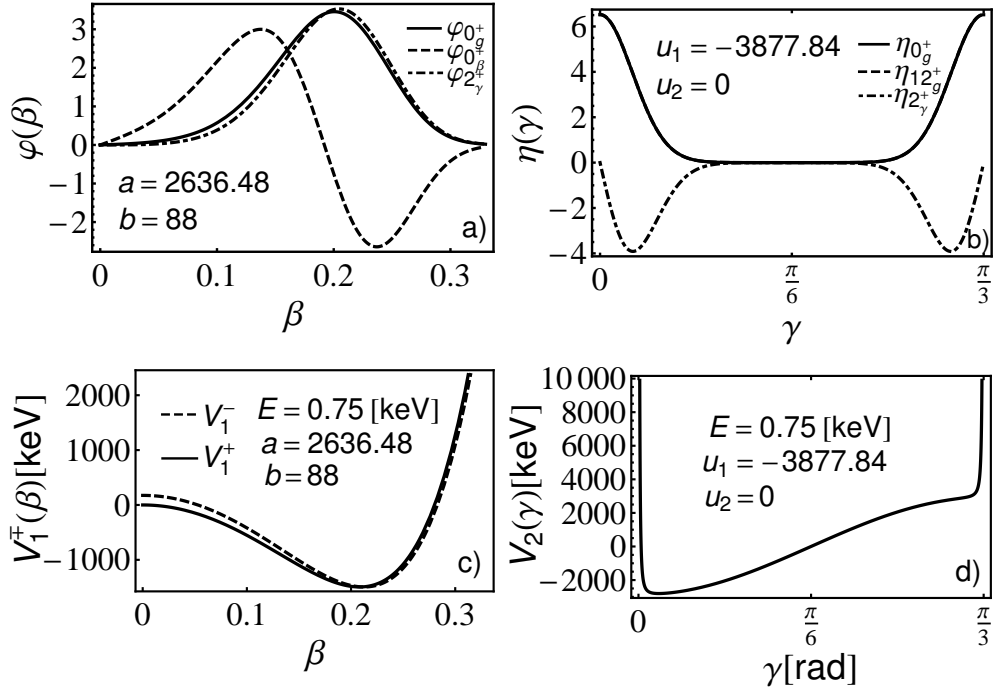}
\end{center}
\caption{The solutions for the equation in $\beta$, corresponding to various angular momenta and  the potential from the left-bottom panel, are plotted, in panel left-up, as a function of $\beta$. Similarly, on the right column the wave functions for $\gamma$ for different angular momenta and the effective potential shown in the right-bottom panel, are plotted as function of $\gamma$. The results correspond to $^{150}$Nd.}
\label{Fig. 2}    
\end{figure}
\begin{figure}[h!]
\begin{center}
\includegraphics[width=0.8\textwidth]{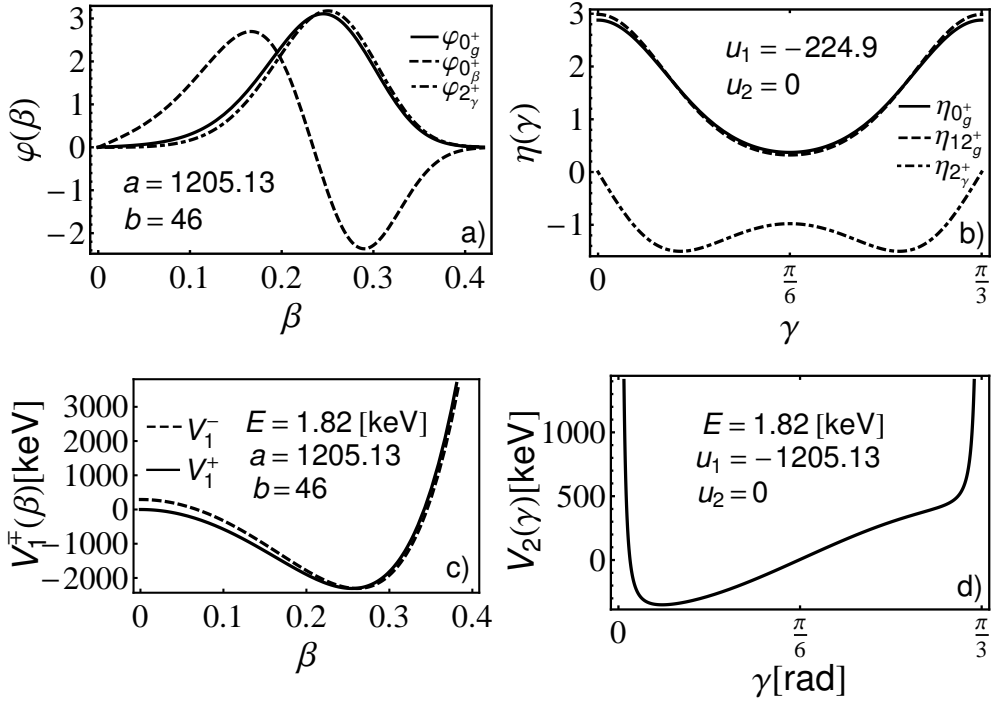}
\end{center}
\caption{The same as in Fig. 2 but for $^{166}$Hf.}    
\label{Fig. 3}
\end{figure}
\begin{figure}[h!]
\begin{center}
\includegraphics[width=0.8\textwidth]{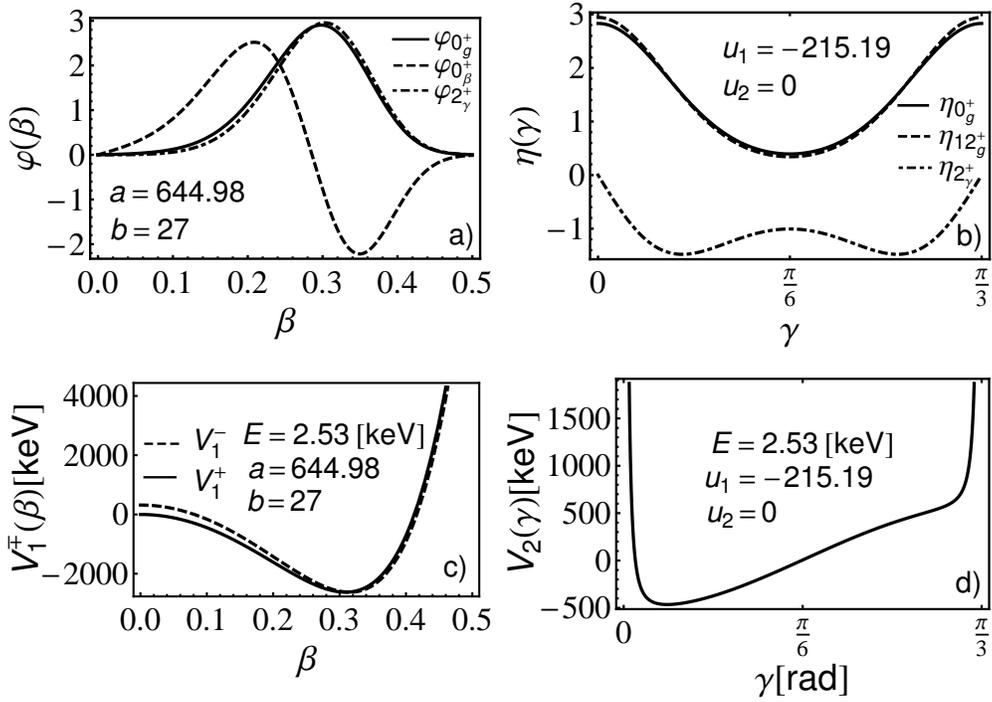}
\end{center}
\caption{The same as in Fig. 2 but for $^{188}$Os.}    
\label{Fig. 4}
\end{figure}
\begin{figure}[h!]
\begin{center}
\includegraphics[width=0.8\textwidth]{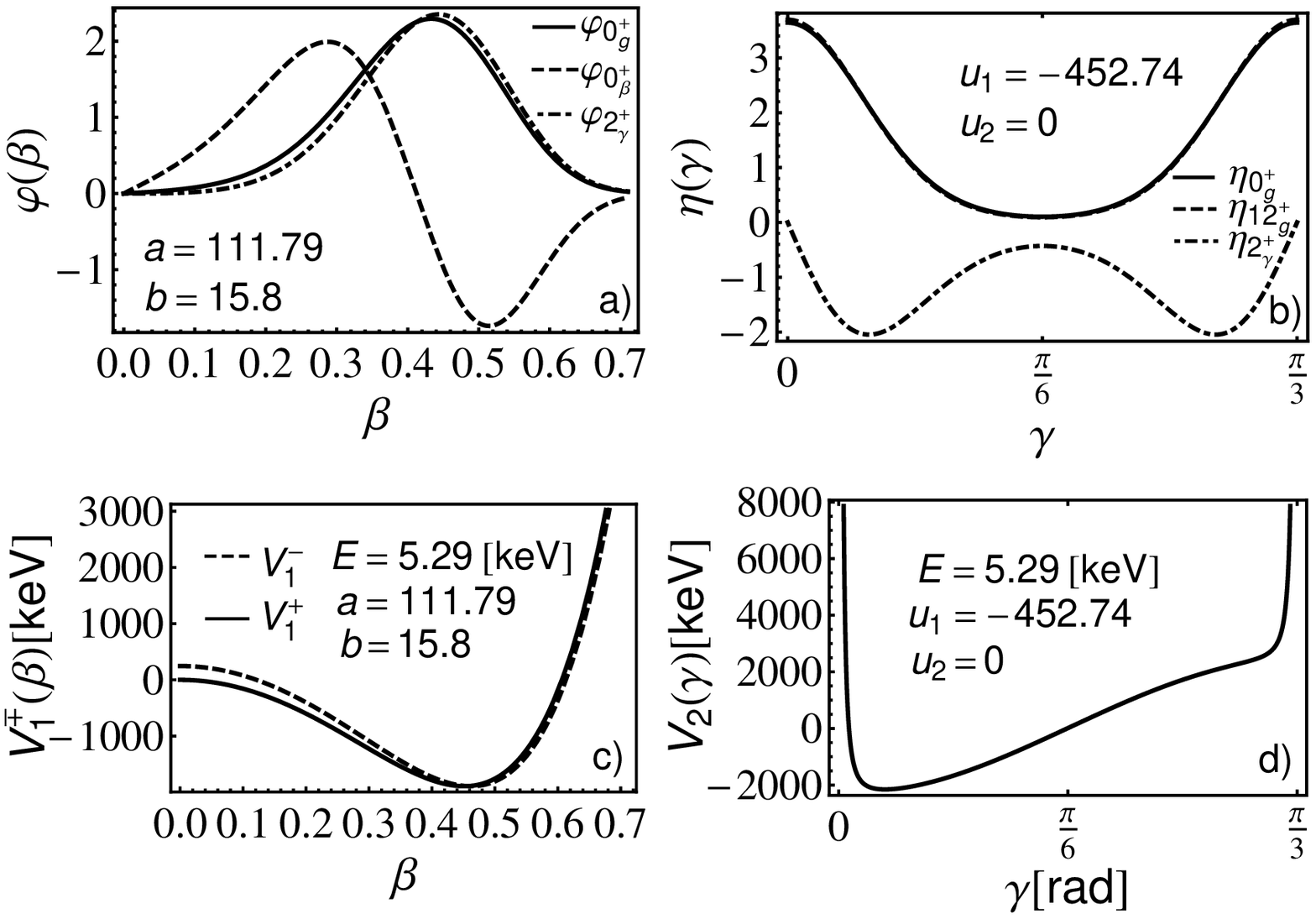}
\end{center}
\caption{The same as in Fig. 2 but for $^{190}$Os.}    
\label{Fig. 5}
\end{figure}
\subsection{Energies}
The  spectra of the chosen nuclei, determined by the models X(5), ISW, D, SSA and CSM, are compared with the corresponding experimental data in Tables VI-XV.
The quality of the agreement between the results of our calculations and the corresponding experimental data is given by the $r.m.s.$ values of the deviations. Thus, comparing the 
$r.m.s.$ values corresponding to different models we conclude that for $^{180}$Os, $^{150}$Nd and
$^{170}$W the best description of the spectra is that given by the CSM approach, energies of
 $^{188}$Os calculated with the $SSA$ are closest to the experimental ones while for the remaining nuclei the $D$ formalism provides the most realist picture.

Using the experimental data listed in Tables VI-XV, one can calculate the ratio of the excitation energies for the states $4^+_g$ and $2^+_g$, denoted by $R_{4^+_g/2^+_g}$. The results are: 2.93 ($^{176}$Os,
$^{190}$Os, $^{150}$Nd, $^{156}$Dy), 2.94 ($^{170}$W), 2.96 ($^{166}$Hf), 3.02 ($^{178}$Os),
 3.08 ($^{188}$Os), 3.10 ($^{180}$Os) and 3.11 ($^{168}$Hf). We notice that all nuclei are characterized by a ratio $R_{4^+_g/2^+_g}$ which is close to the value of 2.9 assigned to the critical point of the transition SU(5)$\to$SU(3), which is described by the solvable model called $X(5)$.
Despite this, the $X(5)$ approach provides a description which is worse than those obtained with the other models proposed here.
 
\begin{table}[h!]
\caption{The energy spectrum of the ground and first $\beta$ and $\gamma$ bands of the $^{176}$Os nucleus yielded by the X(5), ISW, D, SSA and CSM models are compared with the corresponding experimental data taken from Ref. \cite{Basunia}. The energies are given in keV units. The approach which describe best the experimental data is mentioned in a box.}
\vspace{0.3cm}
\resizebox{7.5cm}{10.5cm} {
\begin{tabular}{|c|c|c|c|c|c|c|}
  \hline
  $^{176}$Os & Exp. & X(5) & ISW & $\fbox{D}$ & SSA & CSM \\
  \hline
  $2_{g}^{+}$ & 135 & 126 & 115 & 125 & 125 &135  \\
  $4_{g}^{+}$ & 396 & 367 & 340 & 386 & 377 &394  \\
  $6_{g}^{+}$ & 743 & 686 & 647 & 746& 723  &742  \\
  $8_{g}^{+}$ & 1158 & 1072 & 1026 & 1176 & 1143 &1159  \\
  $10_{g}^{+}$ & 1634 & 1520 & 1473 & 1661 & 1624 &1631  \\
  $12_{g}^{+}$ & 2168 & 2028 & 1986 & 2192 & 2157 &2152  \\
  $14_{g}^{+}$ & 2755 & 2593 & 2564 & 2764 & 2736 &2718  \\
  $16_{g}^{+}$ & 3382 & 3216 & 3205 & 3374 & 3354 &3326  \\
  $18_{g}^{+}$ & 4019 & 3894 & 3909 & 4017 & 4008 &3973  \\
  $20_{g}^{+}$ & 4683 & 4628 & 4673 & 4693 & 4695 &4660  \\
  $22_{g}^{+}$ & 5399 & 5417 & 5499 & 5399 & 5412 &5385  \\
  $24_{g}^{+}$ & 6147 & 6261 & 6385 & 6134 & 6157 &6147  \\
  \hline
  $0_{\beta}^{+}$ & 601 & 714 & 565 & 633 & 498 & 601 \\
  $2_{\beta}^{+}$ & 742 & 942 & 760 & 757 & 723 & 742 \\
  $4_{\beta}^{+}$ & 1026 & 1351 & 1118 & 1019 & 1075 & 1032 \\
  $6_{\beta}^{+}$ & 1432 & 1865 & 1578 & 1378 & 1511 & 1432 \\
  $8_{\beta}^{+}$ &  & 2458 & 2121 & 1808 & 2011 &1914  \\
  $10_{\beta}^{+}$ &  & 3121 & 2738 & 2293 & 2565 &2411  \\
  \hline
  $2_{\gamma}^{+}$ & 864 & 949 & 951 & 926 & 943 & 989 \\
  $3_{\gamma}^{+}$ & 1038 & 1058 & 1056 & 1045 & 1053 &1081  \\
  $4_{\gamma}^{+}$ & 1224 & 1189 & 1184 & 1196 & 1195 & 1201 \\
  $5_{\gamma}^{+}$ & 1410 & 1340 & 1333 & 1371 & 1345 & 1342 \\
  $6_{\gamma}^{+}$ &  & 1509 & 1503 & 1568 & 1542 &1511  \\
  $7_{\gamma}^{+}$ &  & 1694 & 1691 & 1784 & 1719 & 1689 \\
  $8_{\gamma}^{+}$ &  & 1895 & 1898 & 2016 & 1962 & 1900 \\
  $9_{\gamma}^{+}$ &  & 2111 & 2124 & 2264 & 2161 & 2106 \\
  $10_{\gamma}^{+}$ &  & 2343 & 2367 & 2525 & 2444 & 2354 \\
  \hline
  $r.m.s.$ [keV] &  & 156 & 119 & 25 & 41 & 39 \\
  \hline
\end{tabular}}
\end{table}

\begin{table}
\caption{The same as in the Table VI, but for $^{178}$Os. The experimental data are taken from Ref. \cite{Achterberg}.}
\begin{tabular}{|c|c|c|c|c|c|c|}
  \hline
  $^{178}$Os & Exp. & X(5) & ISW & $\fbox{D}$ & SSA & CSM \\
  \hline
  $2_{g}^{+}$ & 132 & 127 & 116 & 131 & 130 &132  \\
  $4_{g}^{+}$ & 399 & 368 & 342 & 402 & 388 & 389 \\
  $6_{g}^{+}$ & 762 & 688 & 650 & 769 & 739 & 736 \\
  $8_{g}^{+}$ & 1194 & 1075 & 1031 & 1203 & 1163 & 1152 \\
  $10_{g}^{+}$ & 1682 & 1525 & 1479 & 1689 & 1647 & 1625 \\
  $12_{g}^{+}$ & 2220 & 2033 & 1994 & 2220 & 2181 &2147  \\
  $14_{g}^{+}$ & 2805 & 2600 & 2572 & 2789 & 2758 & 2715 \\
  $16_{g}^{+}$ & 3429 & 3224 & 3214 & 3395 & 3374 &3325  \\
  $18_{g}^{+}$ & 4020 & 3905 & 3918 & 4033 & 4025 & 3975 \\
  $20_{g}^{+}$ & 4663 & 4641 & 4684 & 4701 & 4706 &4664  \\
  $22_{g}^{+}$ & 5382 & 5432 & 5510 & 5399 & 5415 &5391  \\
  $24_{g}^{+}$ & 6155 & 6278 & 6397 & 6125 & 6150 & 6155 \\
  \hline
  $0_{\beta}^{+}$ & 651 & 716 & 574 & 635 & 493 &651  \\
  $2_{\beta}^{+}$ & 771 & 944 & 771 & 766 & 730 &771  \\
  $4_{\beta}^{+}$ & 1023 & 1355 & 1133 & 1037 & 1092 &1029  \\
  $6_{\beta}^{+}$ & 1396 & 1870 & 1598 & 1403 & 1535 & 1396 \\
  $8_{\beta}^{+}$ &  & 2464 & 2144 & 1838 & 2041 &1850  \\
  $10_{\beta}^{+}$ &  & 3129 & 2766 & 2324 & 2599 &2374  \\
  \hline
  $2_{\gamma}^{+}$ & 864 & 945 & 947 & 916 & 936 &999  \\
  $3_{\gamma}^{+}$ & 1032 & 1055 & 1052 & 1041 & 1048 &1091  \\
  $4_{\gamma}^{+}$ & 1213 & 1187 & 1181 & 1195 & 1195 &1211  \\
  $5_{\gamma}^{+}$ & 1416 & 1338 & 1331 & 1375 & 1346 &1350  \\
  $6_{\gamma}^{+}$ &  & 1507 & 1501 & 1575 & 1546 &1519  \\
  $7_{\gamma}^{+}$ &  & 1692 & 1690 & 1793 & 1725 &1696  \\
  $8_{\gamma}^{+}$ &  & 1894 & 1898 & 2027 & 1971 &1907  \\
  $9_{\gamma}^{+}$ &  & 2111 & 2123 & 2275 & 2170 &2113  \\
  $10_{\gamma}^{+}$ &  & 2343 & 2367 & 2537 & 2455 &2361  \\
  \hline
  $r.m.s.$ [keV] &  & 170 & 141 & 22 & 61 & 54 \\
  \hline
\end{tabular}
\end{table}

\begin{table}
\caption{The same as in the Table VI, but for $^{180}$Os. The experimental data are taken from Ref. \cite{WuNiu}.}
\begin{tabular}{|c|c|c|c|c|c|c|}
  \hline
  $^{180}$Os & Exp. & X(5) & ISW & D & SSA & $\fbox{CSM}$ \\
  \hline
  $2_{g}^{+}$ & 132 & 131 & 124 & 133 & 125 &147  \\
  $4_{g}^{+}$ & 409 & 381 & 374 & 412 & 384 & 423 \\
  $6_{g}^{+}$ & 795 & 713 & 723 & 792 & 748 & 792 \\
  $8_{g}^{+}$ & 1257 & 1115 & 1163 & 1244 & 1196 & 1234 \\
  $10_{g}^{+}$ & 1768 & 1580 & 1688 & 1752 & 1716 &1735  \\
  $12_{g}^{+}$ & 2309 & 2108 & 2297 & 2308 & 2299 & 2291 \\
  $14_{g}^{+}$ & 2875 & 2695 & 2987 & 2906 & 2937 & 2897 \\
  \hline
  $0_{\beta}^{+}$ & 736 & 742 & 522 & 669 & 555 & 736 \\
  $2_{\beta}^{+}$ & 831 & 979 & 720 & 802 & 774 & 831 \\
  $4_{\beta}^{+}$ & 1053 & 1404 & 1093 & 1080 & 1137 & 1051 \\
  $6_{\beta}^{+}$ & 1379 & 1938 & 1584 & 1460 & 1596 & 1379 \\
  $8_{\beta}^{+}$ &  & 2554 & 2175 & 1912 & 2133  & 1799 \\
  $10_{\beta}^{+}$ &  & 3243 & 2858 & 2421 & 2734 & 2299 \\
  \hline
  $2_{\gamma}^{+}$ & 870 & 1011 & 975 & 935 & 985 & 969 \\
  $3_{\gamma}^{+}$ & 1023 & 1125 & 1090 & 1062 & 1100 &1068  \\
  $4_{\gamma}^{+}$ & 1197 & 1262 & 1233 & 1221 & 1245 &1198  \\
  $5_{\gamma}^{+}$ & 1406 & 1418 & 1402 & 1406 & 1402 &1348  \\
  $6_{\gamma}^{+}$ & 1627 & 1593 & 1596 & 1614 & 1609 &1529  \\
  $7_{\gamma}^{+}$ & 1881 & 1786 & 1813 & 1841 & 1797 &1718  \\
  $8_{\gamma}^{+}$ &  & 1995 & 2054 & 2084 & 2057 & 1944 \\
  $9_{\gamma}^{+}$ & 2411 & 2220 & 2318 & 2344 & 2270 & 2164 \\
  $10_{\gamma}^{+}$ &  & 2460 & 2604 & 2617 & 2577 & 2429 \\
  \hline
 $r.m.s.$ [keV] &  & 194 & 96 & 38 & 92 & 35 \\
  \hline
 \end{tabular}
 \end{table}
\begin{table}
\caption{The same as in the Table VI, but for $^{188}$Os. The experimental data are taken from Ref. \cite{Balraj}.}
\begin{tabular}{|c|c|c|c|c|c|c|}
  \hline
  $^{188}$Os & Exp. & X(5) & ISW & D & $\fbox{SSA}$ & CSM \\
  \hline
  $2_{g}^{+}$ & 155 & 179 & 179 & 151 & 152 & 150 \\
  $4_{g}^{+}$ & 478 & 519 & 519 & 479 & 476 & 468 \\
  $6_{g}^{+}$ & 940 & 970 & 970 & 945 & 935 & 934 \\
  $8_{g}^{+}$ & 1515 & 1516 & 1516 & 1512 & 1501 & 1535 \\
  $10_{g}^{+}$ & 2170 & 2149 & 2150 & 2156 & 2154 & 2264 \\
  $12_{g}^{+}$ & 2856 & 2867 & 2868 & 2860 & 2877 & 3116 \\
  \hline
  $0_{\beta}^{+}$ & 1086 & 1009 & 1007 & 1120 & 1063 & 1164 \\
  $2_{\beta}^{+}$ & 1305 & 1331 & 1328 & 1270 & 1330 & 1305 \\
  $4_{\beta}^{+}$ &  & 1910 & 1907 & 1599 & 1808 & 1621 \\
  $6_{\beta}^{+}$ &  & 2636 & 2632 & 2064 & 2421 & 2096 \\
  $8_{\beta}^{+}$ &  & 3474 & 3470 & 2632 & 3132 & 2717 \\
  $10_{\beta}^{+}$ &  & 4412 & 4407 & 3276 & 3920 & 3475 \\
  \hline
  $2_{\gamma}^{+}$ & 633 & 631 & 631 & 627 & 641 & 665 \\
  $3_{\gamma}^{+}$ & 790 & 786 & 785 & 773 & 791 & 790\\
  $4_{\gamma}^{+}$ & 966 & 972 & 971 & 959 & 969 & 956\\
  $5_{\gamma}^{+}$ & 1181 & 1185 & 1185 & 1180 & 1172 & 1157\\
  $6_{\gamma}^{+}$ & 1425 & 1423 & 1423 & 1432 & 1434 & 1399\\
  $7_{\gamma}^{+}$ & 1686 & 1685 & 1684 & 1709 & 1674 & 1669 \\
  $8_{\gamma}^{+}$ &  & 1969 & 1969 & 2009 & 2008 & 1983 \\
  $9_{\gamma}^{+}$ &  & 2275 & 2275 & 2329 & 2273 & 2318\\
  $10_{\gamma}^{+}$ &  & 2602 & 2603 & 2666 & 2670 & 2701\\
  \hline
  $r.m.s.$ [keV] &  & 27 & 27 & 16 & 13 & 36\\
  \hline
\end{tabular}
\end{table}

\begin{table}
\caption{The same as in the Table VI, but for $^{190}$Os. The experimental data are taken from Ref. \cite{Balraj190Os}.}
\begin{tabular}{|c|c|c|c|c|c|c|}
  \hline
  $^{190}$Os & Exp. & X(5) & ISW & $\fbox{D}$ & SSA & CSM \\
  \hline
  $2_{g}^{+}$ & 187 & 188 & 182 & 178 & 172 & 180 \\
  $4_{g}^{+}$ & 548 & 547 & 541 & 551 & 531 & 531 \\
  $6_{g}^{+}$ & 1050 & 1022 & 1034 & 1062 & 1034 & 1031 \\
  $8_{g}^{+}$ & 1666 & 1597 & 1647 & 1672 & 1653 & 1670 \\
  $10_{g}^{+}$ & 2357 & 2264 & 2373 & 2359 & 2367 & 2441 \\
  \hline
  $0_{\beta}^{+}$ & 912 & 1063 & 862 & 925 & 860 & 912 \\
  $2_{\beta}^{+}$ & 1115 & 1402 & 1166 & 1103 & 1168 & 1072 \\
  $4_{\beta}^{+}$ &  & 2012 & 1729 & 1476 & 1682 & 1417 \\
  $6_{\beta}^{+}$ &  & 2777 & 2457 & 1987 & 2331 & 1925 \\
  $8_{\beta}^{+}$ &  & 3659 & 3319 & 2596 & 3083 & 2582 \\
  $10_{\beta}^{+}$ &  & 4647 & 4305 & 3283 & 3921 & 3380 \\
  \hline
  $2_{\gamma}^{+}$ & 558 & 627 & 594 & 583 & 593 & 618 \\
  $3_{\gamma}^{+}$ & 756 & 789 & 756 & 750 & 754 & 756 \\
  $4_{\gamma}^{+}$ & 955 & 985 & 955 & 957 & 954 & 939 \\
  $5_{\gamma}^{+}$ & 1204 & 1210 & 1187 & 1199 & 1172 & 1156 \\
  $6_{\gamma}^{+}$ & 1474 & 1461 & 1451 & 1469 & 1459 & 1419 \\
  $7_{\gamma}^{+}$ &  & 1736 & 1745 & 1764 & 1718 & 1708 \\
  $8_{\gamma}^{+}$ & 2090 & 2035 & 2067 & 2081 & 2080 & 2045 \\
  $9_{\gamma}^{+}$ &  & 2358 & 2419 & 2417 & 2370 & 2401 \\
  $10_{\gamma}^{+}$ & 2772 & 2702 & 2799 & 2770 & 2798 & 2810 \\
  \hline
  $r.m.s.$ [keV] &  & 98 & 26 & 10 & 27 & 36 \\
  \hline
\end{tabular}
\end{table}

\begin{table}
\caption{The same as in the Table VI, but for $^{150}$Nd. The experimental data are taken from Ref. \cite{Dermateosian}.}
\begin{center}
\begin{tabular}{|c|c|c|c|c|c|c|}
  \hline
  $^{150}$Nd & Exp. & X(5) & ISW & D & SSA & $\fbox{CSM}$\\
  \hline
  $2_{g}^{+}$ & 130 & 124 & 121 & 124 & 111 &130  \\
  $4_{g}^{+}$ & 381 & 361 & 358 & 384 & 348 &386  \\
  $6_{g}^{+}$ & 720 & 675 & 682 & 738 & 683 &734  \\
  $8_{g}^{+}$ & 1130 & 1054 & 1084 & 1158 & 1098 &1149  \\
  $10_{g}^{+}$ & 1599 & 1494 & 1560 & 1625 & 1580 &1618  \\
  $12_{g}^{+}$ & 2119 & 1993 & 2106 & 2129 & 2118 &2133  \\
  $14_{g}^{+}$ & 2683 & 2549 & 2722 & 2664 & 2707 &2688  \\
  \hline
  $0_{\beta}^{+}$ & 675 & 702 & 580 & 739 & 630 & 675 \\
  $2_{\beta}^{+}$ & 851 & 926 & 783 & 863 & 822 & 852 \\
  $4_{\beta}^{+}$ & 1138 & 1328 & 1157 & 1123 & 1158 & 1167 \\
  $6_{\beta}^{+}$ & 1541 & 1833 & 1639 & 1477 & 1590 & 1541 \\
  $8_{\beta}^{+}$ &  & 2415 & 2209 & 1897 & 2095 & 1931 \\
  $10_{\beta}^{+}$ &  & 3067 & 2859 & 2364 & 2661 &2319  \\
  \hline
  $2_{\gamma}^{+}$ & 1062 & 1091 & 1087 & 1076 & 1091 &1101  \\
  $3_{\gamma}^{+}$ & 1201 & 1198 & 1197 & 1195 & 1197 &1191  \\
  $4_{\gamma}^{+}$ & 1353 & 1327 & 1333 & 1345 & 1328 &1310  \\
  $5_{\gamma}^{+}$ &  & 1476 & 1491 & 1518 & 1474 &1448  \\
  $6_{\gamma}^{+}$ &  & 1641 & 1671 & 1713 & 1663 &1615  \\
  $7_{\gamma}^{+}$ &  & 1823 & 1872 & 1924 & 1838 & 1790 \\
  $8_{\gamma}^{+}$ &  & 2020 & 2093 & 2151 & 2079 & 1998 \\
  $9_{\gamma}^{+}$ &  & 2233 & 2334 & 2390 & 2276 & 2201 \\
  $10_{\gamma}^{+}$ &  & 2461 & 2594 & 2641 & 2561 & 2445 \\
  \hline
  $r.m.s.$ [keV] &  & 114 & 48 & 28 & 29 & 20 \\
  \hline
\end{tabular}
\end{center}
\end{table}

\begin{table}
\caption{The same as in the Table VI, but for $^{156}$Dy. The experimental data are taken from Ref. \cite{Reich}.}
\begin{tabular}{|c|c|c|c|c|c|c|}
  \hline
  $^{156}$Dy & Exp. & X(5) & ISW & $\fbox{D}$ & SSA & CSM \\
  \hline
  $2_{g}^{+}$ & 138 & 119 & 114 & 140 & 131 &168  \\
  $4_{g}^{+}$ & 404 & 345 & 344 & 422 & 391 &457  \\
  $6_{g}^{+}$ & 770 & 646 & 668 & 796 & 745 &829  \\
  $8_{g}^{+}$ & 1216 & 1009 & 1079 & 1230 & 1175 &1267  \\
  $10_{g}^{+}$ & 1725 & 1431 & 1572 & 1712 & 1667 &1761  \\
  $12_{g}^{+}$ & 2286 & 1908 & 2145 & 2232 & 2151 &2307  \\
  $14_{g}^{+}$ & 2888 & 2440 & 2796 & 2787 & 2807 &2899  \\
  \hline
  $0_{\beta}^{+}$ & 676 & 672 & 451 & 648 & 461 & 676 \\
  $2_{\beta}^{+}$ & 829 & 886 & 629 & 788 & 703 & 829 \\
  $4_{\beta}^{+}$ & 1088 & 1272 & 966 & 1070 & 1068 &1102  \\
  $6_{\beta}^{+}$ & 1437 & 1755 & 1413 & 1444 & 1515 &1452  \\
  $8_{\beta}^{+}$ & 1859 & 2313 & 1955 & 1878 & 2026 &1859  \\
  $10_{\beta}^{+}$ & 2316 & 2937 & 2584 & 2360 & 2593 & 2312 \\
  \hline
  $2_{\gamma}^{+}$ & 891 & 1069 & 898 & 839 & 928 &921  \\
  $3_{\gamma}^{+}$ & 1022 & 1172 & 1004 & 970 & 1041 &1024  \\
  $4_{\gamma}^{+}$ & 1168 & 1296 & 1136 & 1129 & 1188 &1159  \\
  $5_{\gamma}^{+}$ & 1336 & 1438 & 1292 & 1312 & 1339 &1312  \\
  $6_{\gamma}^{+}$ & 1525 & 1596 & 1472 & 1514 & 1542 &1497  \\
  $7_{\gamma}^{+}$ & 1729 & 1771 & 1674 & 1732 & 1720 &1686  \\
  $8_{\gamma}^{+}$ & 1959 & 1960 & 1899 & 1964 & 1972 & 1913 \\
  $9_{\gamma}^{+}$ & 2192 & 2163 & 2145 & 2210 & 2171 & 2131 \\
  $10_{\gamma}^{+}$ & 2448 & 2381 & 2413 & 2467 & 2464 &2395  \\
  $11_{\gamma}^{+}$ & 2712 & 2613 & 2702 & 2735 & 2680 &2636  \\
  $12_{\gamma}^{+}$ & 2997 & 2859 & 3013 & 3013 & 2949 &2934  \\
  $13_{\gamma}^{+}$ & 3274 & 3118 & 3345 & 3301 & 3240 &3153  \\
  $14_{\gamma}^{+}$ &  & 3391 & 3698 & 3600 & 3606 &3526  \\
  $15_{\gamma}^{+}$ & 3861 & 3677 & 4071 & 3908 & 3847 &3805  \\
  \hline
  $r.m.s.$ [keV] &  & 232 & 114 & 35 & 90 &41  \\
  \hline
\end{tabular}
\end{table}

\clearpage

\begin{table}
\caption{The same as in the Table VI, but for $^{166}$Hf. The experimental data are taken from Ref. \cite{Baglin}.}
\begin{tabular}{|c|c|c|c|c|c|c|}
  \hline
  $^{166}$Hf & Exp. & X(5) & ISW & $\fbox{D}$ & SSA & CSM \\
  \hline
  $2_{g}^{+}$ & 159 & 164 & 164 & 152 & 149 &177  \\
  $4_{g}^{+}$ & 470 & 476 & 476 & 471 & 458 &488  \\
  $6_{g}^{+}$ & 897 & 891 & 890 & 906 & 883 &897  \\
  $8_{g}^{+}$ & 1406 & 1391 & 1392 & 1415 & 1392 &1385  \\
  $10_{g}^{+}$ & 1972 & 1973 & 1975 & 1973 & 1966 &1943  \\
  $12_{g}^{+}$ & 2566 & 2631 & 2635 & 2566 & 2588 &2568  \\
  \hline
  $0_{\beta}^{+}$ & 1065 & 926 & 921 & 1064 & 1000 & 1098 \\
  $2_{\beta}^{+}$ & 1219 & 1222 & 1215 & 1216 & 1286 &1219  \\
  $4_{\beta}^{+}$ &  & 1753 & 1745 & 1536 & 1761 & 1490 \\
  $6_{\beta}^{+}$ &  & 2419 & 2410 & 1970 & 2344 &1870  \\
  $8_{\beta}^{+}$ &  & 3189 & 3178 & 2479 & 3002 &2342  \\
  $10_{\beta}^{+}$ &  & 4049 & 4038 & 3038 & 3713 &2893  \\
  \hline
  $2_{\gamma}^{+}$ & 810 & 862 & 862 & 854 & 864 &899  \\
  $3_{\gamma}^{+}$ & 1007 & 1004 & 1003 & 997 & 1007 &1011  \\
  $4_{\gamma}^{+}$ &  & 1174 & 1174 & 1177 & 1178 & 1160 \\
  $5_{\gamma}^{+}$ & 1419 & 1370 & 1370 & 1385 & 1364 & 1330 \\
  $6_{\gamma}^{+}$ &  & 1589 & 1588 & 1617 & 1611 & 1535 \\
  $7_{\gamma}^{+}$ &  & 1829 & 1829 & 1867 & 1822 & 1748 \\
  $8_{\gamma}^{+}$ &  & 2090 & 2090 & 2133 & 2132 & 2002 \\
  $9_{\gamma}^{+}$ &  & 2370 & 2372 & 2411 & 2357 &2251  \\
  $10_{\gamma}^{+}$ &  & 2671 & 2673 & 2701 & 2720 & 2550 \\
  \hline
  $r.m.s.$ [keV] &  & 51 & 53 & 18 & 38 &39  \\
  \hline
\end{tabular}
\end{table}

\begin{table}
\caption{The same as in the Table VI, but for $^{168}$Hf. The experimental data are taken from Ref. \cite{Coral}.}
\begin{tabular}{|c|c|c|c|c|c|c|}
  \hline
  $^{168}$Hf & Exp. & X(5) & ISW & $\fbox{D}$ & SSA & CSM \\
  \hline
  $2_{g}^{+}$ & 124 & 141 & 140 & 120 & 108 &128  \\
  $4_{g}^{+}$ & 386 & 409 & 409 & 382 & 351 &389  \\
  $6_{g}^{+}$ & 757 & 765 & 769 & 757 & 710 &756  \\
  $8_{g}^{+}$ & 1214 & 1195 & 1207 & 1215 & 1172 &1206  \\
  $10_{g}^{+}$ & 1736 & 1694 & 1720 & 1736 & 1723 &1730  \\
  $12_{g}^{+}$ & 2306 & 2259 & 2303 & 2307 & 2354 & 2320 \\
  \hline
  $0_{\beta}^{+}$ & 942 & 795 & 755 & 928 & 878 & 942 \\
  $2_{\beta}^{+}$ & 1059 & 1049 & 1002 & 1048 & 1039 &1049  \\
  $4_{\beta}^{+}$ & 1285 & 1505 & 1449 & 1310 & 1368 &1285  \\
  $6_{\beta}^{+}$ &  & 2077 & 2015 & 1684 & 1823 & 1630 \\
  $8_{\beta}^{+}$ &  & 2738 & 2672 & 2143 & 2380 & 2068 \\
  $10_{\beta}^{+}$ &  & 3477 & 3412 & 2664 & 3024 & 2587 \\
  \hline
  $2_{\gamma}^{+}$ & 876 & 911 & 906 & 902 & 928 & 939 \\
  $3_{\gamma}^{+}$ & 1031 & 1033 & 1028 & 1020 & 1042 & 1035 \\
  $4_{\gamma}^{+}$ & 1161 & 1179 & 1178 & 1172 & 1171 &1161  \\
  $5_{\gamma}^{+}$ & 1386 & 1347 & 1350 & 1353 & 1334 &1311  \\
  $6_{\gamma}^{+}$ & 1551 & 1535 & 1543 & 1558 & 1530 & 1492 \\
  $7_{\gamma}^{+}$ &  & 1741 & 1755 & 1786 & 1733 & 1687 \\
  $8_{\gamma}^{+}$ &  & 1965 & 1988 & 2033 & 1992 & 1916 \\
  $9_{\gamma}^{+}$ &  & 2206 & 2239 & 2297 & 2226 & 2148 \\
  $10_{\gamma}^{+}$ &  & 2464 & 2508 & 2576 & 2543 & 2421 \\
  \hline
  $r.m.s.$ [keV] &  & 75 & 70 & 15 & 43 & 31 \\
  \hline
\end{tabular}
\end{table}

\begin{table}
\caption{The same as in the Table VI, but for $^{170}$W. The experimental data are taken from Ref. \cite{CoralBaglin}.}
\begin{tabular}{|c|c|c|c|c|c|c|}
  \hline
  $^{170}$W & Exp. & X(5) & ISW & D & SSA & $\fbox{CSM}$ \\
  \hline
  $2_{g}^{+}$ & 157 & 145 & 133 & 145 & 151 &171  \\
  $4_{g}^{+}$ & 462 & 420 & 398 & 447 & 446 &475  \\
  $6_{g}^{+}$ & 876 & 785 & 767 & 858 & 844 &873  \\
  $8_{g}^{+}$ & 1363 & 1226 & 1228 & 1347 & 1323 &1346  \\
  $10_{g}^{+}$ & 1902 & 1739 & 1777 & 1894 & 1869 &1882  \\
  $12_{g}^{+}$ & 2464 & 2319 & 2411 & 2489 & 2471 &2477  \\
  $14_{g}^{+}$ & 3118 & 2965 & 3128 & 3126 & 3124 &3128  \\
  $16_{g}^{+}$ & 3816 & 3677 & 3927 & 3801 & 3821 &3831  \\
  \hline
  $0_{\beta}^{+}$ &  & 816 & 587 & 760 & 507 & 823 \\
  $2_{\beta}^{+}$ & 953 & 1077 & 804 & 905 & 790 &953  \\
  $4_{\beta}^{+}$ & 1202 & 1545 & 1208 & 1207 & 1204 &1215  \\
  $6_{\beta}^{+}$ & 1578 & 2132 & 1736 & 1618 & 1706 & 1578 \\
  $8_{\beta}^{+}$ &  & 2810 & 2367 & 2107 & 2277 & 2020 \\
  $10_{\beta}^{+}$ &  & 3568 & 3093 & 2654 & 2905 & 2531 \\
  \hline
  $2_{\gamma}^{+}$ & 937 & 944 & 945 & 928 & 936 & 965 \\
  $3_{\gamma}^{+}$ & 1074 & 1068 & 1068 & 1066 & 1064 &1074  \\
  $4_{\gamma}^{+}$ & 1220 & 1219 & 1219 & 1238 & 1231 & 1217 \\
  $5_{\gamma}^{+}$ &  & 1391 & 1397 & 1438 & 1400 & 1381 \\
  $6_{\gamma}^{+}$ &  & 1584 & 1600 & 1662 & 1630 & 1578 \\
  $7_{\gamma}^{+}$ &  & 1796 & 1828 & 1906 & 1828 & 1783 \\
  $8_{\gamma}^{+}$ &  & 2025 & 2080 & 2168 & 2109 & 2027 \\
  $9_{\gamma}^{+}$ &  & 2273 & 2355 & 2446 & 2329 & 2264 \\
  $10_{\gamma}^{+}$ &  & 2538 & 2653 & 2739 & 2655 & 2550 \\
  \hline
  $r.m.s.$ [keV] &  & 200 & 90 & 21 & 58 & 13 \\
  \hline
\end{tabular}
\end{table}

\subsection{Reduced transition probabilities}

As mentioned before the parameters involved in the transition operators employed by different models have been fixed by fitting through a least square procedure the existent data. With the fitted parameter the results for the reduced E2 transition probabilities are presented in Tables XVI-XXV where one gives for comparison also the available experimental data.
For the lightest three isotopes of $Os$ as well as for $^{166,168}$Hf  and $^{170}$W the available experimental data refers to the states of ground band. The agreements with the experimental data showed up by the five theoretical models are comparable in quality.

For $^{156}Dy$, besides the intraband transitions in the ground band, few interband transitions from the gamma to the ground band are experimentally known. As seen from Table XXII the agreement of calculations  with the experimental data is quite good.

In Ref.\cite{Kruc} measured data in $^{150}$Nd for intraband transitions ground to ground and beta to beta as well interband transitions to ground band have been reported. These data are described reasonably well by the five approaches as shown in Table XXI.  One remarks the good agreement obtained with the CSM approach. The largest
discrepancies with the experimental data are obtained for the transitions $4^+_{\beta}\to 2^+_{g}$ and $4^+_{\gamma}\to 2^+_{g}$ which are overestimated by the theoretical results.

As for $^{188,190}$Os the available data  are about the intraband transitions ground to ground and gamma to gamma bands as well about the interband transition beta to ground and gamma to ground.
They are compared with the results of our calculations in Tables XIX and XX. Again, the agreement
qualities obtained with the five sets of calculations are comparable with each other. The predictions for the decay
probabilities of the transitions $4^+_{\gamma}\to 2^+_{g}$ and $6^+_{\gamma}\to 4^+_{g}$ are  larger than the corresponding experimental data. Also the result for $0^+_{\beta}\to 2^+_{\gamma}$
, obtained within the CSM is about 6.5 larger than the corresponding experimental value.
For some cases the value of the $t_2$ obtained through the least square procedure is very large.
The reason is as follows. 

Within the SSA, the $t_2$ term of the transition operator contribute mainly to the interband transitions while its matrix elements between states of a given band are very small. However, for the mentioned cases there are only few experimental data for interband transitions, most of the data referring to the intraband transitions.  Consequently, the least square procedure is using small matrix elements of the intraband transitions which results in obtaining huge numbers for $t_2$. An equally good description of these cases would be obtained by ignoring the $t_2$ term. We kept however this term just for the sake of having an unitary approach.

The results for the E2 transitions raise the question why the models $X(5), ISW, D, SSA$
predict close results although the states involved are described by different wave functions in the variables $\beta$ and $\gamma$. It seems that these differences are washed out
by the fitting procedure adopted for the strengths of the transition operator . Moreover, the factor function depending on the Euler angles are common in the mentioned 4 approaches, this giving the dominant contribution to the reduced transition probability.

One signature for the triaxiality of the nuclear shape is the equality:
\begin{equation}
E_{2^+_1}+E_{2^+_2}=E_{3^+_1}
\end{equation}
The departure from this rule, $\Delta E=|E_{2^+_1}+E_{2^+_2}-E_{3^+_1}|$, is equal to 2 and 11 keV for $^{188}$Os and $^{190}$Os, respectively. The magnitude of these deviations was the argument for treating the two isotopes as triaxial nuclei \cite{RaBu011}. On the other hand the ratio 
$E_{4/2}$ amounts 2.93 and 3.08 for $^{188}$Os and $^{190}$Os respectively, which are quite close to the specific value of $X(5)$ nuclei. Given these facts we asked ourselves whether these nuclei are axially symmetric or behave like a triaxial rigid rotor. In order to answer this question we compared the r.m.s. values of deviations for both energies and $B(E2)$ values provided by the SMA and SSA approaches, respectively. Concerning the excitation energies in the three major bands, the r.m.s. of prediction deviations from the corresponding experimental data  yielded by the SMA for 
$^{188}$Os and $^{190}$Os are 24 and 32 keV respectively while the SSA  results for these values are 13 and 27 keV. Therefore regarding the excitation energies the two isotopes behave more like  axially deformed nuclei. However comparing the results for the reduced transition probabilities it comes out the triaxial rigid rotor behavior is favored. Indeed, the r.ms.values for the  SMA  approach applied to $^{188}$Os and $^{190}$Os  are 13 W.u. and 16 W.u. respectively, while those corresponding to the SSA, are 16 W. u. and 17 W.u., respectively. Remarkable the fact that the differences of r.m.s values characterizing the two approaches, SMA and SSA are quite small.
Therefore one could conclude that the two investigations, from Ref. \cite{RaBu011} and from here,
indicate that the two nuclei might be equally well described by both approaches. 

\begin{table}
\caption{The reduced E2 transition probabilities determined with the X(5), ISW, D, SSA and CSM models for the $^{176}$Os nucleus are compared with the corresponding experimental data taken from Ref. \cite{Melon}.}
\begin{tabular}{|c|c|c|c|c|c|c|}
  \hline
  B(E2)(W.u.) & Exp. & X(5) & ISW & D & SSA & CSM \\
  \hline
  $2_{g}^{+}\rightarrow0_{g}^{+}$ & $144_{-5}^{+5}$ & 167 & 127 & 145 & 136 &144  \\
  $4_{g}^{+}\rightarrow2_{g}^{+}$ & $243_{-5}^{+5}$ & 264 & 224 & 228 & 227 &253  \\
  $6_{g}^{+}\rightarrow4_{g}^{+}$ & $305_{-11}^{+11}$ & 330 & 305 & 292 & 297 &328  \\
  $8_{g}^{+}\rightarrow6_{g}^{+}$ & $321_{-14}^{+15}$ & 379 & 377 & 360 & 366 &393  \\
  $10_{g}^{+}\rightarrow8_{g}^{+}$ & $441_{-63}^{+88}$ & 419 & 438 & 433 & 435 &452  \\
  $12_{g}^{+}\rightarrow10_{g}^{+}$ & $517_{-146}^{+336}$ & 450 & 490 & 510 & 504 &517  \\
  \hline
\end{tabular}
\end{table}

\begin{table}
\caption{The same as in Table XVI, but for $^{178}$Os. The experimental data are taken from Refs. \cite{Melon,Kibedi,Moller}}
\vspace{0.3cm}
\begin{tabular}{|c|c|c|c|c|c|c|}
  \hline
  B(E2)(W.u.) & Exp. & X(5) & ISW & D & SSA & CSM \\
  \hline
  $2_{g}^{+}\rightarrow0_{g}^{+}$ & 138 & 147 & 137 & 146 & 141 & 138 \\
  \hline
  $4_{g}^{+}\rightarrow2_{g}^{+}$ & 226 & 232 & 225 & 226 & 226 & 227 \\
  \hline
  $6_{g}^{+}\rightarrow4_{g}^{+}$ & 290 & 291 & 287 & 280 & 283 & 282 \\
  \hline
  $8_{g}^{+}\rightarrow6_{g}^{+}$ & 327 & 334 & 337 & 332 & 334 & 327 \\
  \hline
  $10_{g}^{+}\rightarrow8_{g}^{+}$ & 384 & 369 & 378 & 384 & 382 & 368 \\
  \hline
\end{tabular}
\end{table}

\begin{table}
\caption{The same as in Table XVI, but for $^{180}$Os. The experimental data are taken from Ref. \cite{WuNiu}.}
\begin{tabular}{|c|c|c|c|c|c|c|}
  \hline
  B(E2)(W.u.) & Exp. & X(5) & ISW & D & SSA & CSM \\
  \hline
  $2_{g}^{+}\rightarrow0_{g}^{+}$ & $120_{-30}^{+30}$ & 70 & 152 & 148 & 151 & 150 \\
  $4_{g}^{+}\rightarrow2_{g}^{+}$ & $193_{-25}^{+25}$ & 111 & 167 & 177 & 172 & 149 \\
  $6_{g}^{+}\rightarrow4_{g}^{+}$ & $160_{-40}^{+40}$ & 139 & 132 & 139 & 135 & 120 \\
  $8_{g}^{+}\rightarrow6_{g}^{+}$ & $63_{-13}^{+13}$ & 160 & 95 & 83 & 90 & 96 \\
  \hline
\end{tabular}
\end{table}

\begin{table}[h!]
\caption{The same as in Table XVI, but for $^{188}$Os. The experimental data are taken from Ref. \cite{Balraj}. }
\begin{tabular}{|c|c|c|c|c|c|c|}
  \hline
  B(E2)(W.u.) & Exp. & X(5) & ISW & D & SSA & CSM \\
  \hline
  $2_{g}^{+}\rightarrow0_{g}^{+}$ & $79_{-2}^{+2}$ & 74 & 72 & 79 & 82 &42  \\
  $4_{g}^{+}\rightarrow2_{g}^{+}$ & $133_{-8}^{+8}$ & 118 & 115 & 121 & 123 &87  \\
  $6_{g}^{+}\rightarrow4_{g}^{+}$ & $138_{-8}^{+8}$ & 147 & 144 & 147 & 145 & 125 \\
  $8_{g}^{+}\rightarrow6_{g}^{+}$ & $161_{-11}^{+11}$ & 169 & 166 & 174 & 162 &161  \\
  $10_{g}^{+}\rightarrow8_{g}^{+}$ & $188_{-25}^{+25}$ & 187 & 184 & 203 & 178 &195  \\
  \hline
  $0_{\beta}^{+}\rightarrow2_{g}^{+}$ & $0.95_{-0.08}^{+0.08}$ & 47 & 48 & 33 & 21 &0.95  \\
  $0_{\beta}^{+}\rightarrow2_{\gamma}^{+}$ & $4.3_{-0.5}^{+0.5}$ &5.2  & 5.2 & 1.9 & 1.5 &44  \\
  \hline
  $4_{\gamma}^{+}\rightarrow2_{\gamma}^{+}$ & $47_{-10}^{+10}$ & 47 & 50 & 52 & 56 & 14 \\
  $4_{\gamma}^{+}\rightarrow3_{\gamma}^{+}$ & $320_{-120}^{+120}$ & 112 & 117 & 120 & 132 & 43 \\
  $6_{\gamma}^{+}\rightarrow4_{\gamma}^{+}$ & $70_{-30}^{+30}$ & 107 & 111 & 114 & 118 &31  \\
  \hline
  $2_{\gamma}^{+}\rightarrow0_{g}^{+}$ & $5_{-0.6}^{+0.6}$ & 8.4 & 10.9 & 10.8 & 9.9 & 5 \\
  $2_{\gamma}^{+}\rightarrow2_{g}^{+}$ & $16_{-2}^{+2}$ & 13 & 17 & 16 & 14 & 10.4 \\
  $2_{\gamma}^{+}\rightarrow4_{g}^{+}$ & $34_{-5}^{+5}$ & 0.65 & 0.85 & 0.80 & 0.73 & 1.4 \\
  $4_{\gamma}^{+}\rightarrow2_{g}^{+}$ & $1.29_{-0.19}^{+0.19}$ & 5.7 & 7.1 & 6.7 & 6.1 & 1.7 \\
  $4_{\gamma}^{+}\rightarrow4_{g}^{+}$ & $19_{-3}^{+3}$ & 18 & 23 & 20 & 19 & 10.7 \\
  $4_{\gamma}^{+}\rightarrow6_{g}^{+}$ & $16_{-7}^{+7}$ & 2 & 2 & 2 & 2 & 5 \\
  $6_{\gamma}^{+}\rightarrow4_{g}^{+}$ & $0.21_{-0.11}^{+0.11}$ & 5.3 & 6.4 & 5.8 & 5.3 & 0.9 \\
  $6_{\gamma}^{+}\rightarrow6_{g}^{+}$ & $>$9.4 & 21 & 25 & 23 & 20 & 8.3 \\
  \hline
\end{tabular}
\end{table}

\begin{table}
\caption{The same as in Table XVI, but for $^{190}$Os. The experimental data are taken from Ref. \cite{Balraj190Os}.}
\begin{tabular}{|c|c|c|c|c|c|c|}
  \hline
  B(E2)(W.u.) & Exp. & X(5) & ISW & D & SSA & CSM \\
  \hline
  $2_{g}^{+}\rightarrow0_{g}^{+}$ & $72_{-2}^{+2}$ & 58 & 57 & 56 & 61 &45  \\
  $4_{g}^{+}\rightarrow2_{g}^{+}$ & $105_{-6}^{+6}$ & 91 & 91 & 88 & 94 &83  \\
  $6_{g}^{+}\rightarrow4_{g}^{+}$ & $113_{-10}^{+10}$ & 115 & 113 & 112 & 112 &112  \\
  $8_{g}^{+}\rightarrow6_{g}^{+}$ & $137_{-20}^{+20}$ & 131 & 130 & 138 & 126 &137  \\
  $10_{g}^{+}\rightarrow8_{g}^{+}$ & $120_{-30}^{+30}$ & 145 & 143 & 165 & 139 & 160 \\
  \hline
  $0_{\beta}^{+}\rightarrow2_{g}^{+}$ & $2.2_{-0.5}^{+0.5}$ & 36 & 36 & 30 & 19 &2.2  \\
  $0_{\beta}^{+}\rightarrow2_{\gamma}^{+}$ & $23_{-7}^{+7}$ &8.9 & 9 & 8 & 5 & 148 \\
  \hline
  $4_{\gamma}^{+}\rightarrow2_{\gamma}^{+}$ & $53_{-5}^{+5}$ & 36 & 38 & 37 & 41 & 20.4 \\
  $4_{\gamma}^{+}\rightarrow3_{\gamma}^{+}$ & $65_{-13}^{+13}$ & 87 & 90 & 87 & 98 & 84 \\
  $6_{\gamma}^{+}\rightarrow4_{\gamma}^{+}$ & $65_{-13}^{+13}$ & 83 & 85 & 84 & 89 & 49 \\
  $8_{\gamma}^{+}\rightarrow6_{\gamma}^{+}$ & $61_{-16}^{+16}$ & 112 & 113 & 119 & 115 & 72 \\
  \hline
  $2_{\gamma}^{+}\rightarrow0_{g}^{+}$ & $5.9_{-0.6}^{+0.6}$ & 14.2 & 15.6 & 15.9 & 16.2 & 14 \\
  $2_{\gamma}^{+}\rightarrow2_{g}^{+}$ & $33_{-4}^{+4}$ & 21 & 24 & 24 & 24 & 33 \\
  $4_{\gamma}^{+}\rightarrow2_{g}^{+}$ & $0.68_{-0.06}^{+0.06}$ & 9.7 & 10.3 & 10.4 & 10.3 &4.3  \\
  $4_{\gamma}^{+}\rightarrow4_{g}^{+}$ & $30_{-4}^{+4}$ & 31 & 33 & 33 & 32 & 31 \\
  $6_{\gamma}^{+}\rightarrow4_{g}^{+}$ & $<$0.8 & 9 & 10 & 10 & 9 & 1.7 \\
  $6_{\gamma}^{+}\rightarrow6_{g}^{+}$ & $31_{-8}^{+8}$ & 36 & 38 & 40 & 36 &26  \\
  \hline
\end{tabular}
\end{table}

\begin{table}
\caption{The same as in Table XVI, but for $^{150}$Nd. The experimental data are taken from Ref. \cite{Kruc}.}
\begin{tabular}{|c|c|c|c|c|c|c|}
  \hline
  B(E2)(W.u.) & Exp. & X(5) & ISW & D & SSA & CSM \\
  \hline
  $2_{g}^{+}\rightarrow0_{g}^{+}$ & $115_{-2}^{+2}$ & 107 & 104 & 92 & 116 & 81 \\
  $4_{g}^{+}\rightarrow2_{g}^{+}$ & $182_{-2}^{+2}$ & 169 & 168 & 144 & 177 &160  \\
  $6_{g}^{+}\rightarrow4_{g}^{+}$ & $210_{-2}^{+2}$ & 212 & 210 & 183 & 211 &222  \\
  $8_{g}^{+}\rightarrow6_{g}^{+}$ & $278_{-25}^{+25}$ & 243 & 243 & 224 & 240 & 278 \\
  $10_{g}^{+}\rightarrow8_{g}^{+}$ & $204_{-12}^{+12}$ & 269 & 269 & 268 & 266 & 330 \\
  \hline
  $2_{\beta}^{+}\rightarrow0_{\beta}^{+}$ & $114_{-23}^{+23}$ & 85 & 83 & 130 & 86 &116  \\
  $4_{\beta}^{+}\rightarrow2_{\beta}^{+}$ & $170_{-51}^{+51}$ & 128 & 125 & 194 & 144 &165  \\
  \hline
  $0_{\beta}^{+}\rightarrow2_{g}^{+}$ & $39_{-2}^{+2}$ & 67 & 73 & 51 & 37 &41.2  \\
  $2_{\beta}^{+}\rightarrow0_{g}^{+}$ & $1.2_{-0.2}^{+0.2}$ & 2.1 & 2.9 & 3.1 & 1.6 &5.2  \\
  $2_{\beta}^{+}\rightarrow2_{g}^{+}$ & $9_{-2}^{+2}$ & 10 & 10 & 9 & 6 & 9 \\
  $2_{\beta}^{+}\rightarrow4_{g}^{+}$ & $17_{-3}^{+3}$ & 39 & 42 & 40 & 26 &26  \\
  $4_{\beta}^{+}\rightarrow2_{g}^{+}$ & $0.12_{-0.02}^{+0.02}$ & 1.07 & 1.61 & 1.64 & 0.57 & 5.6 \\
  $4_{\beta}^{+}\rightarrow4_{g}^{+}$ & $7_{-1}^{+1}$ & 6 & 8 & 8 & 5 & 7.2 \\
  $4_{\beta}^{+}\rightarrow6_{g}^{+}$ & $70_{-13}^{+13}$ & 30 & 33 & 46 & 26 & 26 \\
  \hline
  $2_{\gamma}^{+}\rightarrow0_{g}^{+}$ & $3_{-0.8}^{+0.8}$ & 2.4 & 8 & 9.8 & 5.1 &16.3  \\
  $2_{\gamma}^{+}\rightarrow2_{g}^{+}$ & $5.4_{-1.7}^{+1.7}$ & 3.6 & 11.9 & 14.3 & 7.3 &5.4  \\
  $2_{\gamma}^{+}\rightarrow4_{g}^{+}$ & $2.6_{-2.0}^{+2.0}$ & 0.2 & 0.6 & 0.7 & 0.4 & 0.74 \\
  $4_{\gamma}^{+}\rightarrow2_{g}^{+}$ & $0.9_{-0.3}^{+0.3}$ & 1.6 & 5 & 6.1 & 3 & 28.6 \\
  $4_{\gamma}^{+}\rightarrow4_{g}^{+}$ & $3.9_{-1.2}^{+1.2}$ & 5.3 & 15.5 & 18.9 & 9 & 9.6 \\
  \hline
\end{tabular}
\end{table}

\clearpage

\begin{table}
\caption{The same as in Table XVI, but for $^{156}$Dy. The experimental data are taken from Ref. \cite{Reich}. }
\begin{tabular}{|c|c|c|c|c|c|c|}
  \hline
  B(E2)(W.u.) & Exp. & X(5) & ISW & D & SSA & CSM \\
  \hline
  $2_{g}^{+}\rightarrow0_{g}^{+}$ & $149.3_{-2.5}^{+2.5}$ & 142 & 138 & 111 & 137 &66  \\
  $4_{g}^{+}\rightarrow2_{g}^{+}$ & $261_{-17}^{+17}$ & 225 & 223 & 179 & 219 & 149 \\
  $6_{g}^{+}\rightarrow4_{g}^{+}$ & $200_{-15}^{+15}$ & 282 & 279 & 235 & 271 &221  \\
  $8_{g}^{+}\rightarrow6_{g}^{+}$ & $289_{-14}^{+14}$ & 323 & 323 & 295 & 316 & 289 \\
  $10_{g}^{+}\rightarrow8_{g}^{+}$ & $366_{-25}^{+25}$ & 358 & 358 & 359 & 357 & 354 \\
  $12_{g}^{+}\rightarrow10_{g}^{+}$ & $382_{-22}^{+22}$ & 385 & 386 & 425 & 395 & 418 \\
  \hline
  $2_{\gamma}^{+}\rightarrow0_{g}^{+}$ & $7.2_{-0.4}^{+0.4}$ & 6.6 & 9.9 & 23.3 & 11.6 &7.2  \\
  $2_{\gamma}^{+}\rightarrow2_{g}^{+}$ & $9.4_{-1.0}^{+1.0}$ & 9.8 & 14.6 & 35.1 & 17.4 &9.4  \\
  $2_{\gamma}^{+}\rightarrow4_{g}^{+}$ & $12.6_{-1.9}^{+1.9}$ & 0.5 & 0.7 & 1.8 & 0.9 &19.5  \\
  \hline
\end{tabular}
\end{table}

\begin{table}
\caption{The same as in Table XVI, but for $^{166}$Hf. The experimental data are taken from Ref. \cite{Baglin}.}
\begin{tabular}{|c|c|c|c|c|c|c|}
  \hline
  B(E2)(W.u.) & Exp. & X(5) & ISW & D & SSA & CSM \\
  \hline
  $2_{g}^{+}\rightarrow0_{g}^{+}$ & $128_{-7}^{+7}$ & 98 & 153 & 154 & 155 & 128 \\
  $4_{g}^{+}\rightarrow2_{g}^{+}$ & $202_{-7}^{+7}$ & 155 & 212 & 216 & 215 & 203 \\
  $6_{g}^{+}\rightarrow4_{g}^{+}$ & $221_{-13}^{+13}$ & 194 & 225 & 232 & 226 &245  \\
  $8_{g}^{+}\rightarrow6_{g}^{+}$ & $280_{-30}^{+30}$ & 223 & 225 & 230 & 225 &280  \\
  $10_{g}^{+}\rightarrow8_{g}^{+}$ & $250_{-110}^{+640}$ & 246 & 220 & 219 & 218 &311  \\
  $12_{g}^{+}\rightarrow10_{g}^{+}$ & $155_{-70}^{+550}$ & 265 & 213 & 199 & 209 &351  \\
  \hline
\end{tabular}
\end{table}

\begin{table}
\caption{The same as in Table XVI, but for $^{168}$Hf. The experimental data are taken from Ref. \cite{Coral}.}
\begin{tabular}{|c|c|c|c|c|c|c|}
  \hline
  B(E2)(W.u.) & Exp. & X(5) & ISW & D & SSA & CSM \\
  \hline
  $2_{g}^{+}\rightarrow0_{g}^{+}$ & $154_{-7}^{+7}$ & 141 & 165 & 176 & 175 & 154 \\
  $4_{g}^{+}\rightarrow2_{g}^{+}$ & $244_{-12}^{+12}$ & 223 & 250 & 257 & 255 &249  \\
  $6_{g}^{+}\rightarrow4_{g}^{+}$ & $285_{-18}^{+18}$ & 279 & 294 & 292 & 291 &304  \\
  $8_{g}^{+}\rightarrow6_{g}^{+}$ & $350_{-50}^{+50}$ & 320 & 322 & 318 & 316 &350  \\
  $10_{g}^{+}\rightarrow8_{g}^{+}$ & $370_{-60}^{+60}$ & 354 & 342 & 338 & 338 &391  \\
  $12_{g}^{+}\rightarrow10_{g}^{+}$ & $320_{-120}^{+120}$ & 381 & 356 & 354 & 357 &438  \\
  \hline
\end{tabular}
\end{table}

\begin{table}
\caption{The same as in Table XVI, but for $^{170}$W. The experimental data are taken from Ref. \cite{CoralBaglin}.}
\begin{tabular}{|c|c|c|c|c|c|c|}
  \hline
  B(E2)(W.u.) & Exp. & X(5) & ISW & D & SSA & CSM \\
  \hline
  $2_{g}^{+}\rightarrow0_{g}^{+}$ & $124_{-3}^{+3}$ & 79 & 133 & 126 & 129 & 124 \\
  $4_{g}^{+}\rightarrow2_{g}^{+}$ & $179_{-18}^{+18}$ & 125 & 179 & 177 & 179 &168  \\
  $6_{g}^{+}\rightarrow4_{g}^{+}$ & $189_{-14}^{+14}$ & 157 & 184 & 189 & 187 &182  \\
  $8_{g}^{+}\rightarrow6_{g}^{+}$ & $190_{-50}^{+50}$ & 180 & 180 & 187 & 183 &190  \\
  $10_{g}^{+}\rightarrow8_{g}^{+}$ & $170_{-40}^{+40}$ & 199 & 173 & 175 & 174 &197  \\
  $12_{g}^{+}\rightarrow10_{g}^{+}$ & $160_{-30}^{+30}$ & 214 & 167 & 158 & 162 &214  \\
  \hline
\end{tabular}
\end{table}

\clearpage

\section{Conclusions}
Here, we summarize the main results obtained by this work. We selected 10 nuclei characterized by a ratio $R_{4^+_g/2^+_g}$ close to 2.9 which is specific to the so called X(5) nuclei. Spectra of these nuclei are described by a new approach which treats the beta variable by the Schr\"{o}dinger equation associated to a sextic oscillator plus a centrifugal potential. For the variable $\gamma$ one finds a differential equation which is satisfied by the spheroidal function.
The excitation energies are obtained by summing the  eigenvalues provided by the differential equations for the $\beta$ and $\gamma$ variable respectively, while the corresponding functions are used to calculate the E2 transition probabilities. The results are compared with the corresponding experimental data as well as with those obtained through other formalisms called 
X(5), ISW, D and CSM which were earlier used by the present authors  to describe the spectroscopic properties of other X(5) like nuclei. 

Note that while the formalisms X(5), ISW, D and SSA treat the energies and transition probabilities using the intrinsic coordinates and the rotation matrix function, the CSM is a quadrupole boson approach and therefore the mentioned observables are calculated with the collective coordinates which are specific to the laboratory frame.

A comparison of the r.m.s. values yielded by the five approaches shows that the D, CSM and SSA approaches produce the best agreement with the experimental energies. Concerning the
E2 transitions one may say that all five sets of results quantitatively describe the experimental situation in a comparable manner with a slight advantage for SSA and CSM.
Since the formalisms ISW, D, SSA, differ from each other by the way the variable beta is treated, otherwise the $\gamma$ equation being  the same, the transition probabilities produced by the
three approaches exhibit similar agreement with the experimental data. The SSA method produces
very good agreement with the experimental energies for $^{188}$Os, $^{150}$Nd and $^{168}$Hf.
Table V shows that these nuclei have the largest deformations and moreover for the first two nuclei
the ratio $R_{4^+/2^+}$ has the values 3.08 and 3.11 respectively, which deviate most from the 
X(5) value. The quoted ratio for $^{150}$Nd is 2.93 which is close to the X(5) value but its  deformation is the largest one.

The sextic potential for the $\beta$ assures a more realistic description of the excited states where the the excitation of the beta degree of freedom is important. This is best seen in the excellent agreement of the calculated excitation energies in the beta and gamma bands with the 
corresponding experimental data.

The final conclusion is that the SSA, proposed in this paper, proves to be a suitable tool for a realistic description of the X(5) like nuclei.

{\bf Acknowledgment.} This work was supported by the Romanian Ministry for Education Research Youth and Sport through the CNCSIS project ID-2/5.10.2011.

\end{document}